\documentclass[11pt,letterpaper,reprint,notitlepage,superscriptaddress,aps,physrev]{revtex4-2}

\usepackage{amsmath,amssymb,amsfonts} 
\usepackage{empheq}
\usepackage{physics}
\usepackage{multirow}
\usepackage{appendix}
\usepackage{enumitem}
\usepackage{relsize}
\usepackage{pifont}

\setlist{nosep}

\usepackage{graphicx}

\usepackage{subdepth}   

\usepackage{bm} 
\renewcommand{\mathbf}{\bm}
\usepackage{dsfont} 
\renewcommand{\mathbb}{\mathds}

\usepackage[svgnames,dvipsnames]{xcolor}
\usepackage{tikz}
\usepackage{graphicx}
\definecolor{NewBlue}{rgb}{0.1, 0.1, 0.7}
\definecolor{NewRed}{rgb}{0.7, 0.1, 0.1}
\usepackage[colorlinks,
    linkcolor=Maroon,
    citecolor=NewBlue,
    urlcolor=NewRed]{hyperref}

\usepackage{cleveref}

\usepackage{amsthm}
\usepackage{ragged2e}
\newtheorem{defn}{Definition}

\newcommand*\circled[1]{\textbf{\tikz[baseline=(char.base)]{
            \node[shape=circle,fill,inner sep=0.5pt] (char) {\textcolor{white}{#1}};}}}

\newcommand{\RleMIT}{Research Laboratory of Electronics, Massachusetts Institute of Technology, Cambridge, Massachusetts 02139, USA}
\newcommand{\Delft}{QuTech, Delft University of Technology, Lorentzweg 1, 2628 CJ Delft, The Netherlands}

\newcommand{\xmark}{\ding{55}}%

\begin{document}

\title{Scalable Quantum Networks:\\
Congestion-Free Hierarchical Entanglement Routing with Error Correction}

\author{Hyeongrak Choi}
\email{choihr@mit.edu}
\affiliation{\RleMIT}
\author{Marc G. Davis}
\affiliation{\RleMIT}
\author{Álvaro G. Iñesta}
\affiliation{\Delft}
\author{Dirk R. Englund}
\email{englund@mit.edu}
\affiliation{\RleMIT}

\date{\today}

\begin{abstract}
We introduce Quantum Tree Networks (QTN), an architecture for hierarchical multi-flow entanglement routing. 
The network design is a $k$-ary tree where end nodes are situated on the leaves and routers at the internal nodes, with each node connected to $k$ nodes in the child layer. 
The channel length between nodes grows with a rate $a_k$, increasing as one ascends from the leaf to the root node. 
This construction allows for congestion-free and error-corrected operation with qubit-per-node overhead to scale sublinearly with the number of end nodes, $N$.
The overhead for a $k$-ary QTN scales as $\mathcal{O}(N^{\log_k a_k} \cdot \log_k N)$ and is sublinear for all $k$ with minimal surface-covering end nodes. 
More specifically, the overhead of quarternary ($k=4$) QTN is $\sim \mathcal{O}(N^{0.25}\cdot\log_4 N)$. 
Alternatively, when end nodes are distributed over a square lattice, the quaternary tree routing gives the overhead $\sim \mathcal{O}(\sqrt{N}\cdot\log_4 N)$.
Our network-level simulations demonstrate a size-independent threshold behavior of QTNs.
Moreover, tree network routing avoids the necessity for intricate multi-path finding algorithms, streamlining the network operation.
With these properties, the QTN architecture satisfies crucial requirements for scalable quantum networks.
\end{abstract}

\maketitle
\section{Introduction}

Quantum networks (QNs) present a solution to the challenge of transmitting quantum states across space, time, and physical modalities~\cite{cirac1997quantum,briegel1998quantum,wehner2018quantum,ruf2021quantum}. QNs entangle remote quantum memories via optical channels to enable applications such as communication, resource sharing, blind quantum computing~\cite{arrighi2006blind}, and distributed sensing~\cite{Zhang2021Distributed}. A central research goal focuses on developing architectures and algorithms for \emph{entanglement routing}, which consists of distributing entanglement among multiple end nodes with high fidelity and rate~\cite{hahn2019quantum, hayashi2007quantum, pant2019routing, patil2022entanglement, liu2022quantum}.

Photonic channels enable long-distance entanglement due to fast signal propagation, low loss of 0.14 dB/km~\cite{hasegawa2018first}, and negligible thermal noise at room temperature~\cite{rudolph2017optimistic}. Optically heralded entanglement generation maps channel losses into reduced entanglement generation rates without lowering fidelities~\cite{briegel1998quantum,duan2001long,barrett2005efficient,bernien2013heralded}. However, photon-based schemes encounter a significant challenge in long-distance transmission due to the exponential decay of photon transmission. Consequently, the development of quantum repeaters becomes imperative to mitigate photon losses and ensure high generation rates~\cite{briegel1998quantum}. Moreover, within the network, quantum repeaters play an additional role as routers, efficiently distributing entanglements to end nodes that are not directly connected~\cite{pant2019routing}.

To distribute entanglement to two dedicated end nodes, QNs generate short-distance entanglement and perform entanglement swapping over the path connecting the two. The intermediate nodes need to perform the following tasks:
\begin{enumerate}
\item Perform a repeat-until-success procedure of heralded entanglement generation \cite{duan2001long, barrett2005efficient, bernien2013heralded}.
\item Store quantum states in memories until receiving a classical signal of successful entanglement generation.
\item Select two qubits entangled with different nodes and perform Bell measurements for entanglement swapping \cite{briegel1998quantum}.
\item Restore fidelities diminished by imperfections using entanglement purification \cite{bennett1996purification} or quantum error correction~\cite{jiang2009quantum, munro2012quantum}.
\end{enumerate}

The processes require multiple two-way classical communications, particularly for probabilistic results such as entanglement heralding and purification. During the associated time delays, quantum networks either reserve a path or dynamically search for the path between target end nodes. Thus, the network operation is close to the circuit-switched network~\cite{roberts1978evolution}, where the resource usage of a pair blocks the use by others. 

In addition, current and near-future technologies for quantum networks are \emph{memory-limited}; the number of memories is many orders of magnitude smaller than the capacity of optical channels. The state-of-the-art demonstration has shown the use of four memories~\cite{pompli2021}, and even the largest quantum computer without networking capabilities has 433 qubits (IBM Osprey)~\cite{chow2021ibm}. On the other hand, optical channels can transmit millions of single photons within the coherence time of these memories~\cite{zahidy2023quantum}. This is a crucial difference from classical networks, where large memories are readily available, and communication is constrained by channel capacity (\emph{channel-limited} system).

Circuit switching in a memory-constrained network can lead to congestion. This arises when the number of available qubits in a router is exceeded by the simultaneous demands of entanglement routing paths. A trivial solution is to linearly increase the number of memories. The central question is: \emph{What is the minimum number of qubits required for multi-flow entanglement routing across end nodes?}

``\emph{Scalable}'' quantum networks should support multi-flow entanglement routing~\cite{chakraborty2020entanglement}, scaling with the network size, defined as the number of end nodes. This results in a linearly increased aggregated network capacity, maintaining constant throughput. The accumulation of operational errors needs to be prevented for reliable entanglement sharing. Routers must also have sufficient buffers to ensure they are free of congestion. While it is straightforward to achieve all these with a linearly increasing number of qubits per end node — which we term ``overhead'' — the goal is to design a network that has sublinear overhead but retains its efficacy.

\begin{defn}
An $N$-user quantum network is scalable if the network allows: \\
Condition 1: $\mathcal{O}(N)$ error-corrected entanglement flow \\
Condition 2: without congestion \\
Condition 3: with $o(N)$ quantum memories per end node\\
Assumption: under uniform random traffic.
\end{defn}

Here, we assume uniform random traffic in which every end node requests entanglement with every other end node with an equal probability, but the gross probability that one node engages in entanglement is assumed to be fixed (Assumption). The average number of entanglement flow in the network increases linearly with the number of network end nodes (Condition 1). A network is `congestion-free' if the multi-flow routing paths do not exceed the number of memories at the nodes. Formally, we consider a set of requests in a snapshot of networks, that is, a member of the typical set given by the traffic model. In this limit, the probability of communication failure decays exponentially as the network size increases (Condition 2). Lastly, the number of memories per end node is required to be sublinear in a scalable network (Condition 3).

In this work, we propose a `quantum tree network' (QTN) that meets all the conditions necessary for a scalable network design. A QTN only allows for optimal routing paths, enabling fast communication and excluding time-consuming multipath-finding algorithms. We find that the overhead of QTNs can be decomposed into contributions from the repeating-routing and the error correction. If the channel is dominated by insertion loss, the overhead is logarithmic. In the propagation loss dominant regime, the overhead depends on the geometric deployment of the network. If the network grows by adding more end nodes in a fixed area, the overhead remains logarithmic. If the geometric length of the network grows slower than the degree of the tree, the overhead is guaranteed to be sublinear. Specifically for general 2D surface covering, the overhead has a sub-square-root end-node scaling. 

\section{Quantum Network Topology} 

The arrangement of the network resources, such as nodes and edges, defines a network topology and highly affects the performance of a classical network. A broad range of topologies has been studied and deployed. These topologies span from simple configurations like ring, mesh, bus, and star topologies, to complex data processing structures, such as Clos networks, Omega networks, and fat trees.
However, the discussion of quantum network topology has been mostly limited to the 1D repeater chain \cite{muralidharan2016optimal}, the star topology \cite{vardoyan2019stochastic}, and the two-dimensional uniform lattices~\cite{choi2019percolation, pant2019routing, patil2022entanglement}.

\begin{figure*}
    \centering
    \includegraphics[width=\textwidth]{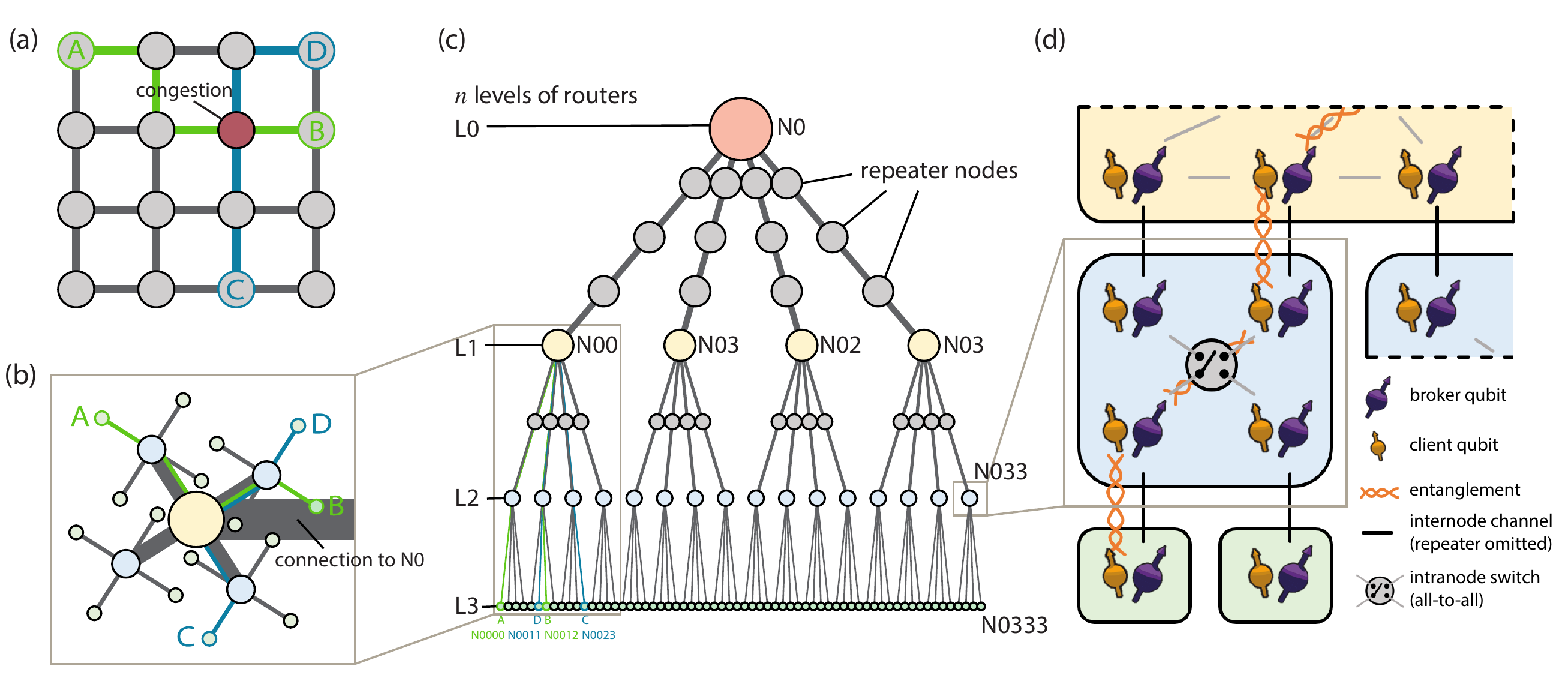}
    \caption{Congestion-free quantum tree network. (a) Quantum mesh network with nodes on a square lattice. Light and dark green lines indicate the routing paths for the A-B and C-D pairs, respectively. The red node is part of both paths and is thus congested. (b) Quantum tree network. The size of the nodes is proportional to the number of qubits they hold, and the width of the edges is proportional to allocated qubit pairs. (c) Hierarchical representation of a quantum tree network ($k=4$). (d) Quantum router architecture with all-to-all intra-node connectivity.}
    \label{fig:TreeNetwork}
\end{figure*}

Although lattice-based quantum mesh networks are robust under local failures and enable distance-independent communication rates~\cite{choi2019percolation,pant2019routing,patil2022entanglement}, memory congestion prevents them from being scalable. Figure~\ref{fig:TreeNetwork}(a) shows a square-lattice quantum mesh network. Each node, represented with a circle, is an end node requesting entanglement. The edges are physical channels through which entanglement is generated. Because every node is only connected to a few neighbors, entanglement must be routed with entanglement swapping. Thus, in a mesh network, each node is a router as well as an end node. Assume that two pairs of end nodes, A-B (light green) and C-D (dark green), request entanglement. The network routes entanglement through one of the shortest paths. If each node has only two qubits, each node can perform only one entanglement swapping at a time. When two routing paths of the requests cross, the network is under congestion. 

What is the total number of qubits for multi-flow routing in a quantum mesh network? Under uniform random traffic, $N_e$ entanglement paths over mesh networks have $\mathcal{O}(N_e^2)$ node intersections (see Appendix~\ref{apdx:CongestionMesh}). For $N_e \sim \mathcal{O}(N)$, the total number of qubits required is $\mathcal{O}(N^2)$. The overhead, the number of qubits per user, is $\mathcal{O}(N)$ in the congestion-free limit. Note that the actual number of qubits used is $\mathcal{O}(N\sqrt{N})$ because the average length of routing paths is $\mathcal{O}(\sqrt{N})$. One does not know where congestion occurs a priori. Thus, all nodes that can be potentially under congestion should be equipped with qubits capable of handling maximum traffic, though in most instances they are not (see Appendix~\ref{apdx:CongestionMesh}). Another way to avoid congestion is through multipath optimization with global information from the network. We assume that this is not available due to computational complexities and finite coherence time (see Appendix~\ref{apdx:multipath} for a detailed discussion). 

In the following discussion, we derive a large number of qubits per node, a requirement that may seem challenging to meet with near-future devices. However, we emphasize that the number of physical qubits can be scaled down through time-multiplexing in practical settings. For instance, a node with 10 qubits and another with just one qubit can be mapped to two nodes, each having a single qubit; one node continuously operates its qubit for entanglement distribution, while the other does at a 10\% duty cycle; during 90\% of remaining bandwidth, the qubit can be allocated for other tasks, e.g. computation. To mitigate the memory occupation of duty-cycle mismatched qubits, one can employ quasi-asynchronous entanglement generation~\cite{munro2010quantum}. In brief, the overhead we present should be understood as the space-time cost of qubit usage required to operate the network.

\section{Quantum Tree Network}

We propose a quantum tree network (QTN) as a scalable quantum network architecture. The QTN architecture employs hierarchical aggregation of routing paths through routers with a corresponding number of qubits. A similar classical network architecture, the fat tree network, is used in data centers, where stable throughput is required for random traffic under a memory constraint~\cite{leiserson1985fat}. Ref. \cite{al2008scalable} showed the scalability and out-performance of classical fat trees compared to other hierarchical designs. 

Figure \ref{fig:TreeNetwork}(b) describes the entanglement routing for the A-B and C-D pairs. The router where two paths intersect has more qubits as illustrated by a larger node to avoid congestion. In the QTN framework, end nodes do not route entanglement. Instead, we employ dedicated routers to route entanglement flows from end nodes and lower-level routers.

Figure \ref{fig:TreeNetwork}(c) details the hierarchical structure of QTN. For a $k$-ary tree and $N$ end nodes, the height of the tree is $n = \log_k N$. The layers with depth-$i$ nodes are labeled L$i$, $i\in\{0,1,...,n\}$. All the nodes but the root and leaves are physically connected to $k$ other nodes in L$(i+1)$ and one node in L$(i-1)$. The nodes are labeled with ``N'' followed by a $k$-ary number, labeling the $x_j^\text{th}$ child attached to the parent N$x_0x_1...x_{j-1}$ as N$x_0x_1...x_{j-1}x_j$. For example, we show A and B in N0000 and N0012. Because they are in different branches of N00, they need entanglement swapping through nodes N000, N00, N001, and N012. Because the tree graph is free of cycles, there is a unique selection of quantum routers for a pair of end nodes. We exemplify our discussion with uniformly branching trees, but our discussion holds for other trees.

Using the notation, Appendix~\ref{apdx:ActProb} derived that routers in an upper layer need to route exponentially more entanglement as $\sim k^{(n-i)}$. Thus, routers in a higher layer should have exponentially more qubits to avoid congestion. Furthermore. the nodes can be buffered $m$ times for faster entanglement generation. All in all, a router in the L$i$ layer has $2mk^{(n-i)}$ qubits with a factor of 2 for the entanglement swapping. 

While the architecture described determines the routing, we need to detail the deployment of the QTN nodes because the distances between the router nodes translate to the number of repeaters and ultimately to the resource overhead. For this, we use the growth rate $a_k$, the ratio of the distance between a router in L$i$ and in L$(i+1)$ to the one between L$(i-1)$ and L$i$. An interesting choice of $a_k$ is the deployment of QTN for minimal 2D surface coverage. Appendix~\ref{apdx:2DCovering} shows that this construction gives $a_k < \sqrt{k}$, and it also details how the router nodes can be deployed to cover nonuniform demands. Another choice of $a^\text{sq}_4 = 2$ is for the end nodes distributed on a square lattice. The channel length between L$i$ and L$(i-1)$ is $\propto a_k^{(n-i)}$, and the number of repeater stations is proportional to the channel length. A repeater between L$i$ and L$(i-1)$ has $2mk^{(n-i)}$ qubits to keep the generation rate proportional to the number of qubits in the nodes. 

Figure~\ref{fig:TreeNetwork} (d) illustrates the ``quantum router architecture'' proposed in ref.~\cite{lee2022quantum} ($m=1$ for visual clarity). We use broker-client qubits for a robust and scalable entanglement distribution~\cite{benjamin2006brokered,choi2019percolation}. Broker qubits (purple) in a node generate heralded entanglement (orange) through the intranode (gray) and internode (black) channels. If the entanglement is heralded, it is stored in client qubits (yellow) with a longer storage time and subsequent entanglement generation. We assume all-to-all intranode connectivities that can possibly be implemented with trapped-ion systems~\cite{monroe2014large,brown2016co} or color centers in diamond with integrated photonics~\cite{wan2020large} or spatial light modulators~\cite{li2022scalable}. We assume that the intranode photon loss is negligible compared to the internode and ignore the temporal resources for the intranode routing.

\section{Resource Overhead Calculation}\label{sec:overhead}

Here, we analyze the resource overhead of the deployed quantum tree networks. We start with a raw entanglement generation rate for each end node, $R = p_e / t_0$, where $p_e$ is the probability of heralding entanglement, and $t_0$ is the trial time. The distance between the nodes in L$i-1$ and those in L$i$ is denoted by $l_0 \cdot a_k^{(n-i)}$, and $l_0$ represents the elementary distance between an end node and the first router node. We normalize the rates with $R$ and the distances with $l_0$, and thereby decoupling them from the calculation. 

If the memory-photon interfaces are inefficient, or if a network grows by adding more end nodes in a finite area, the routers' insertion loss can dominate the exponential channel losses. In this case, which we call the dense-node limit, one does not need repeater stations. If operational errors are small enough for the target applications, the total number of qubits across the network is trivial, $\mathcal{N} \approx 2N\log_k N$. Thus, the overhead scales as $\mathcal{N}/N \approx 2\log_k N$.

If the network is large and errors accumulate over the target error rate of an application, routers should employ quantum error correction~\cite{jiang2009quantum}. The code encodes $n_\text{phys}$ qubits into a smaller number of logical qubits $k_\text{logic}$ and can correct up to $t$ errors ([[$n_\text{phys}, k_\text{logic}, 2t+1$]] code). Thus, $n_\text{phys}/k_\text{logic}$ times more qubits are required to keep the logical entanglement generation rate at the same level as the raw generation rate $R$. We consider a subset of CSS codes called the "good quantum code"~\cite{calderbank1996good} with $n_\text{phys}/k_\text{logic} \lesssim 19t$, saturating the Gilbert-Varsharov bound~\cite{nielsen2002quantum}. The encoding exponent, the power of $t$ in $n_\text{phys}/k_\text{logic}$, of this code is $J=1$ resulting in a desired scaling. However, other codes can be more suitable for practical considerations such as noise models, connectivities, and device dimensionalities. For this reason, we also consider the surface code with $n_\text{phys}/k_\text{logic} = (2t+1)^2$ ($J = 2$), which has been demonstrated to have a high error threshold under circuit noises only with weight-4 stabilizers. (We also analyzed nested error correction using both fixed and variable values for $t$. For those interested, see Appendix~\ref{apdx:resource_overhead}). 

For a given physical error rate $\epsilon$, the logical error rate of entanglement swapping is, 
\begin{align}
    \epsilon_L \approx (\epsilon/\epsilon_\text{th})^t \cdot \epsilon,    
\end{align}
where $\epsilon_\text{th}$ is the error correction threshold. The expression is based on the statistical argument~\cite{fowler2012surface} and is applicable to topological codes under circuit-level noises. However, any quantum error correction codes satisfying the threshold theorem~\cite{nielsen2002quantum} will yield similar scaling. Depending on the code and the operations such as decoding, the equation may vary by a factor of constant, which we assume to be one for the simplicity of discussion. We also exemplify and simplify our calculations with the assumption, $\epsilon/\epsilon_\text{th} = 0.1$, based on current state-of-the-art hardware~\cite{arute2019quantum}.

For a dense-node quantum network, the maximum number of entanglement swapping is $N_\text{swap} = 2\log_k N$. For a given target error $\epsilon_0$,
\begin{align}
    \epsilon_0 &\approx N_\text{swap} \epsilon_L = 2\log_k N 10^{-t} \epsilon.
\end{align}
Assuming $\epsilon_0 = \epsilon$,
\begin{align}
    t \approx \log_{10} (2\log_k N).
\end{align}
Thus, the overhead of error-corrected dense-node network is $\mathcal{N}_\text{CSS}/N \approx 38\log_{10}(2\log_k N)\cdot\log_k N$ and $\mathcal{N}_\text{S.C.}/N \approx 8\left[\log_{10}(2\log_k N)\right]^2\cdot\log_k N$ for CSS and surface-code encoding, respectively.

On the other hand, if the channel loss is dominated by the propagation loss of photons, which we call sparse-node quantum network, we need to consider the additional cost of repeaters. The number of repeaters depends on the geometric details of the deployment. Considering channel length growth rate $a_k$ in the previous section, the number of entanglement swap is,
\begin{align}
    N_\text{swap} = 2\frac{a_k^n-1}{a_k-1}.
\end{align}
Thus, with the same assumption on $\epsilon$, $\epsilon_\text{th}$, and $\epsilon_0$ as above,
\begin{align}
    t \approx n\log_{10}a_k, \label{eq:tForak}
\end{align}
where we assumed $n\log_{10} a_k \gg \log_{10} [2/(a_k-1)]$ and $a_k^n \gg 1$. As a result, the overhead is,
\begin{align}
    &\mathcal{N}/N \approx C_J\cdot N^{\log_k a_k} \cdot (\log_k N)^J, \label{eq:overheadQTN}\\
    &
    \begin{cases}
    J=1, C_1 = 38\cdot \frac{\log_{10}a_k}{a_k-1} & \text{(CSS)}\\
    J=2, C_2 = 8\cdot \frac{[\log_{10}a_k]^2}{a_k-1} & \text{(Surface Code)}.
    \end{cases}
\end{align}
Here, we used the fact that every router layer and repeater layer has the same number of memories, $2N$, except the root and end nodes with $N$, and we multiplied the error correction overhead with the number of repeater and router layers times 2. Thus, we can decompose the overhead into the routing and repeating overhead, $\sim N^{\log_k a_k}$, and the error-correction overhead, $\sim [\log_k N]^J$. (We also analyzed the case separately encoding each router-router channel, which may help modular operations. For those interested, see Appendix~\ref{apdx:router-router_encoding}).

All in all, Eq.~(\ref{eq:overheadQTN}) shows that the minimal overhead scaling is with $J=1$, and $\mathcal{N}/N \sim \mathcal{O} \left(N^{\log_k a_k} \cdot \log_k N\right)$, and the overhead is sublinear for $a_k<k$. This is the case for the minimal surface covering ($a^\text{2D}_k < \sqrt{k}$) and the square-lattice covering ($a^\text{sq}_4 = 2$) as shown in Appendix~\ref{apdx:2DCovering}. For example, a quaternary QTN with minimal surface covering shows the sub-square-root scaling overhead, $\mathcal{N}/N\sim \mathcal{O}(N^{1/4}\log_4 N)$. Similarly, a quarternary QTN embedded on a square lattice has $\mathcal{N}/N\sim \mathcal{O}(\sqrt{N}\log_4 N)$.

\begin{figure*}[ht]
    \centering
    \includegraphics[width=\textwidth]{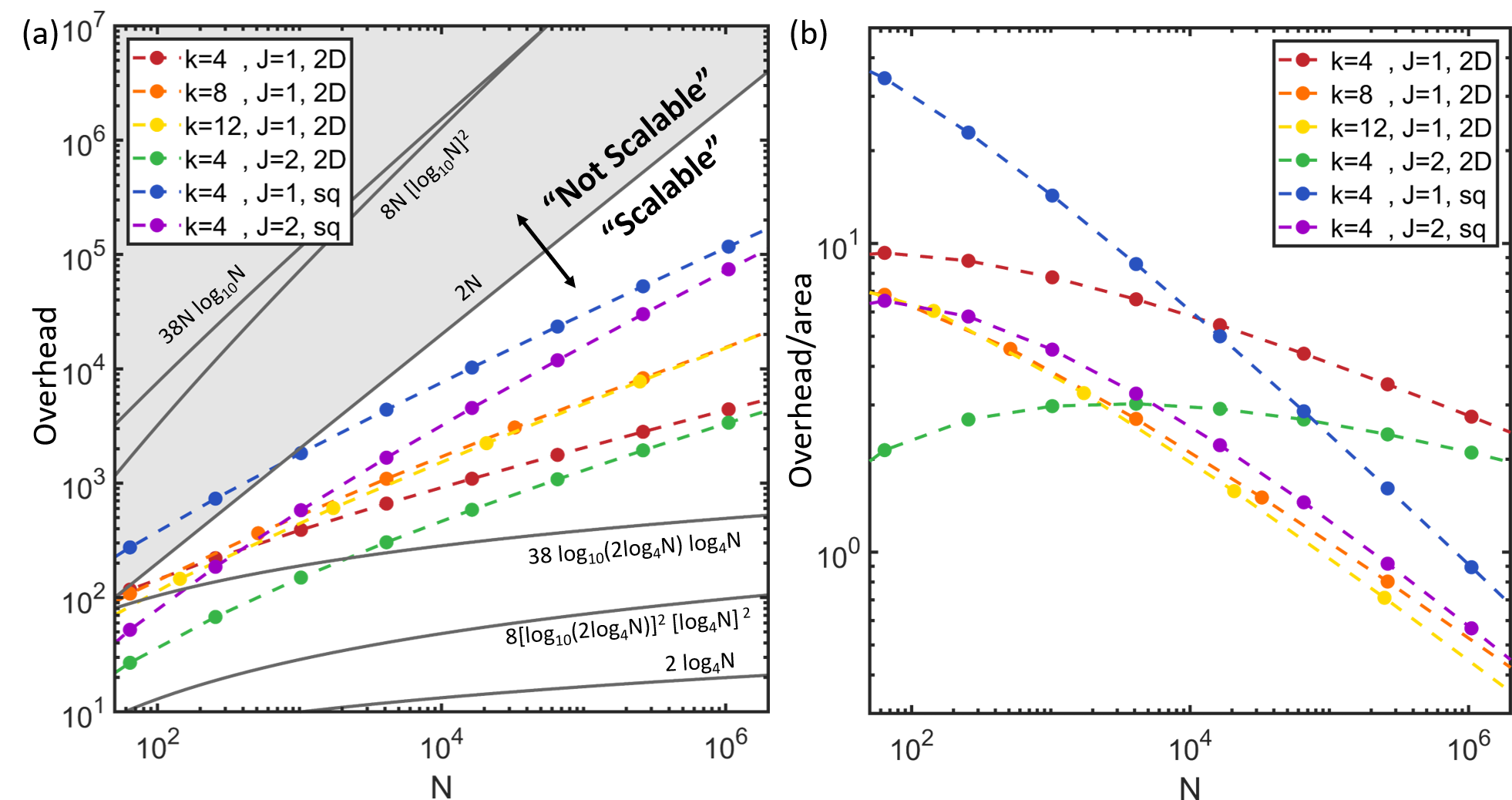}
    \caption{Resource overhead of QTNs for a range of user number $N$, from $4^3$ to $4^{10}$. (a) Equation~(\ref{eq:overheadQTN}) is used for QTN overheads. $J=1$: CSS code encoding, $J=2$: surface code encoding, 2D: minimal surface covering, sq: square-lattice embedding. We plot the gray lines for comparison. $y=2N$: boundary of scalable and non-scalable regimes. $y = 2\log_4 N$: non-error-corrected dense-node quantum network with $k=4$. $y=38N\log_{10}N$, $8N\log_{10}N$: linear routing overhead, with $J=1$ and $J=2$ error correction. $y=38\log_{10}(2\log_4N)\log_4N$, $8\left[\log_{10}(2\log_4N)\right]^2\left[\log_4N\right]^2$: dense-node quantum network with error correction. (b) The area-normalized overheads of QTNs. We approximate the area of the network as $\pi (a_k^nl_0)^2$ and $2(a_k^nl_0)^2$ for minimal surface covering and square lattice embedding.
    }
    \label{fig:resource_overhead}
\end{figure*}

Figure~\ref{fig:resource_overhead}(a) plots the resource overhead of QTNs for the user number $N$ from $4^3$ to $4^{10}$. The red, orange, and yellow markers show the overhead of $k=4, 8, 12$, CSS-coded ($J=1$) QTNs deployed to minimally cover the 2D surface (labeled as ``2D''). The overhead with surface code encoding ($J=2$) increases faster (green) but is smaller for $N \leq 10^6$. If the network is embedded on a square-lattice (blue and purple, labeled ``sq''), the overhead is greater and the scaling is faster than those of the surface-covering case. 

We plot the gray lines for comparison. The line $y=2N$ (the factor of 2 comes from the need for entanglement swapping) separates the scalable and non-scalable regimes. $y = 2\log_4 N$ line shows non-error-corrected and dense-node QNs for $k=4$. Except for a few, all data points are in the scalable regime between the two lines. $y=38N\log_{10}N$ and $8N[\log_{10}N]^2$ lines show linear routing overhead, $N$, combined with the CSS and surface code error correction, respectively, for a fair comparison with error-corrected QTNs. For $N\approx 10^6$, there are five orders of magnitude overhead improvement with QTNs (green) than the error-corrected, linear overhead routings. $y=38\log_{10}(2\log_4N)\log_4N$ and $8\left[\log_{10}(2\log_4N)\right]^2\left[\log_4N\right]^2$ lines plot the overhead for dense-node QTN with the two kinds of error correction. Note that all overheads with $J=1$ and $J=2$ are above the two, respectively. At $k=4$, QTNs have only $\sim 10$ (2D, red) and $\sim 40$ (2D, green) times more overhead than the minimum, up to $N\approx 10^6$. 

Figure~\ref{fig:resource_overhead}(b) plots the area-normalized overheads in units of qubits/end node/$l_0^2$ where $l_0$ is the length of the elementary link. We approximate the area of QTN as $\pi (a_k^nl_0)^2$ and $2(a_k^nl_0)^2$ for minimal surface covering and square-lattice embedding, respectively. Larger $k$-value QTNs (orange and yellow) have a better scaling in $N$ than the small $k$ QTN (red). Similarly, the square lattice QTN (blue and purple) scales better than the surface covering QTN (red and green). Small $k$, surface-covering QTNs are better for dense nodes --- still in sparse-node network limit, where propagation loss dominates the channel loss --- while large $k$, square lattice QTNs are better for a sparse nodes. For a given distribution of end nodes, one can easily optimize the QTNs with nonuniform, heterogeneous branching factors, deployment, and encoding.

It is worth noting that a graph-theoretic study of hierarchical networks \cite{bapat2018unitary} has demonstrated advantages in state transfer and the preparation of a large GHZ state. Moreover, Eldredge et al. found that hierarchical networks outperform conventional networks in creating network-size-scalable bipartite entanglement (``rainbow state'') \cite{eldredge2020entanglement}. In contrast to these seminal works, we focus on the overhead derived from arbitrary targeted entanglement distribution, error correction, network deployment, and the cost of repeaters, as well as congestion in dynamic networks.

\section{Network Simulation of QTN}

In the previous Section, we derived the scaling of the overhead in the number of end nodes with uniform random traffic. We did not explicitly consider the balance of the rate at which the end nodes request entanglement and the entanglement generation rate. Any network can fail if the entanglement request rate is too high. In this Section, we simulate a QTN to show that the network is congestion-free below a threshold request rate, and it starts to fail fulfilling the requests after such a threshold. We also analyze the dynamic response of the network to sudden changes in the request rate.

We perform a network-level simulation of the proposed QTN architecture. We developed a discrete-event simulator in {\tt Python}~\cite{Davis2023QTN}. The simulation performs a sequence of time cycles of duration $t_0$, during which each node attempts entanglement generation. We assume that a successfully generated entanglement lasts 1000 cycles, considering that nitrogen-vacancy (NV) color centers in diamond have a 10~$\mu$s entanglement trial cycle and an electronic coherence time of $10$ ms~\cite{humphreys2018deterministic, pompli2021}. In principle, one can extend the coherence time exponentially with error correction. Alternatively, the state can be stored in $^{13}$C nuclear spins with a minute-scale coherence time~\cite{bradley2019ten}. For simplicity and network-level study agnostic to physical hardware, we abstract these details and treat entanglement binarily; it only lasts within a coherence time of 1000 cycles and expires regardless of fidelities.

We simulate a quarternary ($k=4$) QTN, where each node is buffered with $m=10$ to support a high rate as well as batched communication. The routers communicate with the rate boosted by the repeater chains, which is aggregated as the uniform probability of entanglement success between routers. In each cycle, random pairs of clients request entanglement distribution. There are $n_0 = \binom{N}{2} = N(N-1)/2$ possible communication pairs, each of which randomly requests entanglement with probability $p_0=Np/2n_0$ at each cycle. On average, $Np/2$ requests are created at each cycle, and the probability that a client is in one of the pairs requested is $\approx p$. Thus, we call $p$ the request rate. In the simulation, we sweep $p_0$ and plot network properties from $p = 10^{-3}$ to $p = 10^{-1}$. We also simulate the network under the batched requests of entanglement with the batch size $b$. Each communication pair has a probability $p_0=Np/2n_0b$ of a request at each cycle. Although the requests are batched, each request is still completed individually. 

A request is fulfilled when all the entanglements along the path between the clients are available. Completing a request depletes them, leaving the memories available for a new entanglement generation. We assume that the time for entanglement swapping is negligible compared to that of entanglement generation. Each node attempts entanglement with the dedicated neighboring nodes (Fig.~\ref{fig:TreeNetwork}(d)), with a success probability of $p_e=0.001$ per trial. Once a qubit is successfully entangled, it remains until the entanglement is consumed, or until it expires after 1000 cycles. When the entanglement expires, the qubits are reset and available for new entanglement. The requests are timed out after 1000 cycles as well. The simulation is repeated and averaged with dynamic sample sizes. At every $p$, the simulation was repeated 1000 times or until the 90\% confidence interval was smaller than 1\% of the range of possible values, whichever comes first (10 minimum sampling). 

In the simulation, we measure the success rate, mean latency, and buffered entanglement. The success rate is the ratio of the number of successfully processed requests to that of all requests. The mean latency is the average cycle of requests. Expired requests contribute to the mean latency with the time-out of 1000 cycles. The buffered entanglement is the number of entangled qubit pairs  between the router nodes in the network. 

\begin{figure*}
    \centering
    \includegraphics[width=\textwidth]{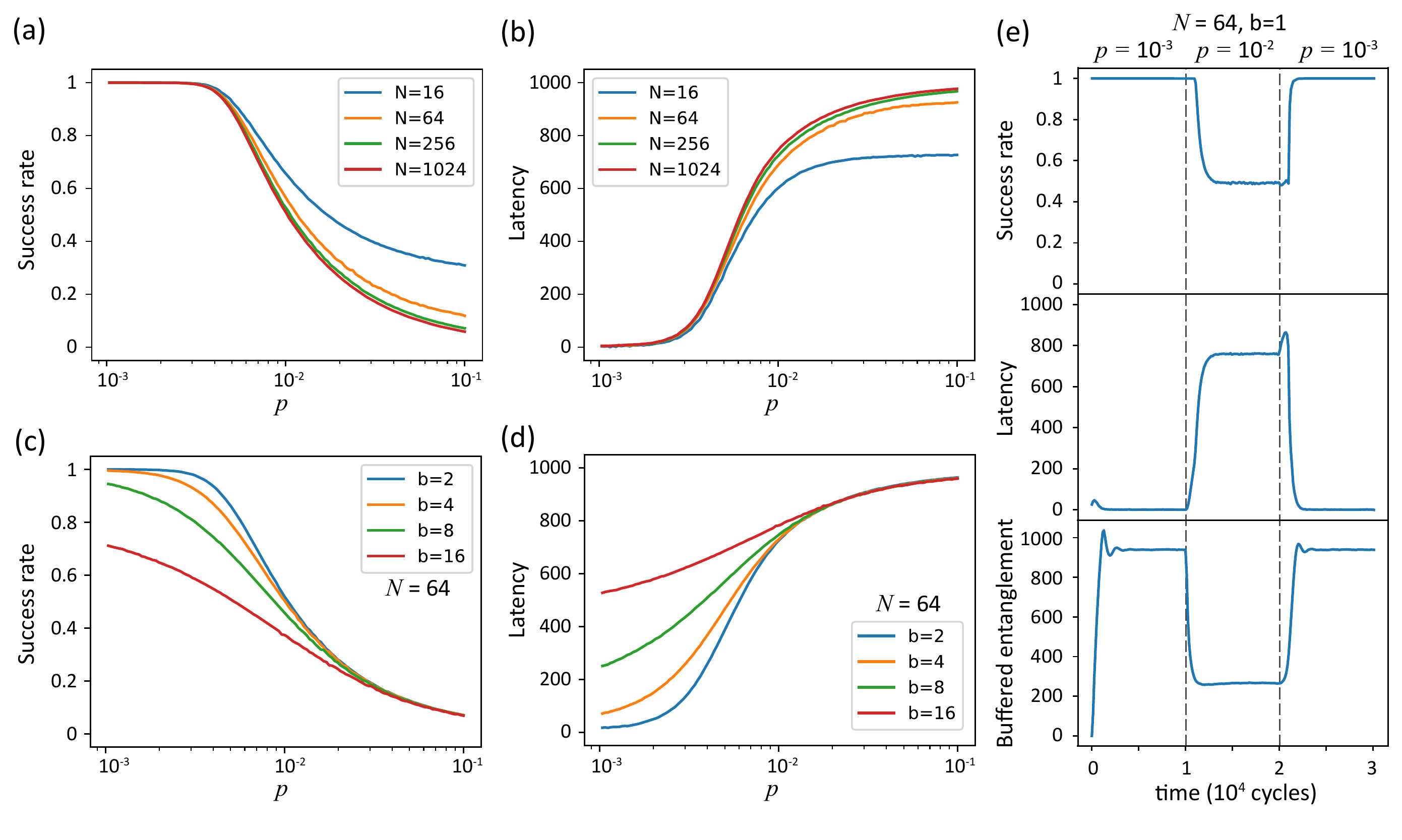}
    \caption{Characterization of the quantum tree network. (a) The success rate of entanglement delivery and (b) mean latency vs. $p$ for different network sizes of $N = 4, 64, 256, 1024$. (c,d) The success rate and mean latency of batched entanglement requests with batch sizes of $b$ = 2, 4, 8, 16 ($N$ = 64). The thresholding is smoothed for large $b$. (a)-(d) The simulation was run repeatedly until the estimate of the mean converged for each data point (see main text). The 90\% confidence region of the estimates of the true mean is shaded but not visible due to small errors. (e) The dynamic responses of QTN. The success rate, the mean latency, and the buffered entanglement of a network ($N=64$) under a temporary increase in $p$ from $10^{-3}$ to $10^{-2}$ between time cycles of $10^4$ and $2\times 10^4$. 512 simulations were averaged in the ensemble and the values were time-averaged with 64-cycle bins.}
    \label{fig:simulations_unified}
\end{figure*}

Figures~\ref{fig:simulations_unified} (a) and (b) show the success rate and mean latency of requests at a rate from $p = 10^{-3}$ to $p = 10^{-1}$ for quaternary QTNs with $N = 16, 64, 256, 1024$. We observe the thresholding behavior of the success rate and the mean latency at $p_\text{th} \approx 0.003$. For $p\lesssim p_\text{th}$, the success rate approaches unity and the mean latency is negligible. For $p\gtrsim p_\text{th}$, the network shows increased mean latency, and a fraction of requests expire. It is important that $p_\text{th}$ is independent of $N$. Above the threshold, both metrics strongly depend on $N$ for small $N$, but converge to an asymptotic limit for large $N$. In the QTN, entanglement flows are bundled as we move up the network hierarchy. This bundling with the routers and repeaters operating first-generated, first-out basis --- in analogy with `first-in first-out' (FIFO) --- relaxes the coherence time requirement under probabilistic entanglement generation~\cite{lee2022quantum} resulting in the near unity success probability for $p<p_\text{th}$.

Figure~\ref{fig:simulations_unified} (c) and (d) plot the success rate and mean latency of the batched requests. We fixed a network of size $N=64$ and varied the size of the batch $b$. For $b=2$, the thresholding behavior remains with a reduced $p_\text{th} \approx 0.002$. Below the threshold, almost all requests were completed with negligible latencies. The higher the value of $b$, the lower the success rates and the higher the mean latencies. Therefore, the curves are smoothed for a large $b$. For $b > m$, the complete usage of buffered entanglement cannot fulfill the batched requests, and the success rate does not reach near unity for small $p$.

Figure~\ref{fig:simulations_unified} (e) shows the dynamic response of QTN ($N=64$) to the change in request rate ($b=1$). The rate changes from $p=10^{-3}$ to $p=10^{-2}$ in time cycle $10^{4}$ and reverses back at time cycle $2\times10^{4}$. These two values are below and above the threshold, respectively. We repeated the simulation $512$ times for averaging and time averaging with discrete time bins of 64 cycles.

After the request rate increased from $0.001$ to $0.01$ at $t=10^4$, the 90\% to 10\% fall-time was approximately 1024 cycles for the success rate and 640 cycles for the buffered entanglement. The 10\% to 90\% rising time in latency was $\sim$1408 cycles. These changes were slightly offset; the drop in memory buffer came first, followed by a rise in latency, and eventually a fall in success rate as the memory buffer was unable to form enough new entanglements to meet demands. Once the request rate returned to $10^{-3}$ at $t=2\cdot10^4$, the buffered entanglement immediately began to recover, with a 10\% to 90\% rising time of approximately 896 cycles. Latency showed a brief spike due to the completion of accumulated requests, which would have expired over the next few cycles if the network remained busy. During this brief period, the success rate remains low. Once the backlog was processed, latency quickly dropped with a 90\% to 10\% falling time of 640 cycles, and the success rate abruptly rose, with a 10\% to 90\% rising time of 256 cycles. Note that these times are measured to the nearest time bucket, which are multiples of 64 cycles. The smaller spike at the beginning of the simulation is because the system starts without any entanglement.

The buffered entanglement exhibited oscillations following sharp increases. This is due to a feedback between entanglement generation and consumption. When there are enough buffered entanglements in the network, it is more likely that the entanglements required for any given request will be already present in the buffer. This causes entanglement resources to be consumed faster when the buffer is filled. Meanwhile, entanglement generation attempts occur only between memories that are not already entangled. Thus, entanglement generation rates are higher when the buffer is empty. This results in the observed oscillation until the system reaches an equilibrium in which the entanglement generation rate equals the consumption rate from request fulfillment (combined with small expiration).

\section{Conclusion}

We introduced the scalability of quantum networks and showed that hierarchical entanglement routing on a quantum tree network satisfies the condition for the scalability. The overhead of QTNs is logarithmic of the network size if the gate error is negligible and the channel loss is independent of layers. The overhead of QTNs with lossy and erroneous channels depends on the deployment. We proposed two scenarios: minimal surface covering and square-lattice embedding. Incorporating second-generation repeaters for router-to-router routing, we showed that the overhead is $\mathcal{O}\left(N^{\log_k a_k}\log_k N\right)$. Furthermore, the quantum tree network does not need time-consuming multipath-finding problems~\cite{li2023swapping} (see also Appendix~\ref{apdx:multipath}) or dynamic routing~\cite{chen2023q}. Our approach is suitable for near-term hardware, including atom-like emitters in diamond~\cite{ruf2021quantum} or trapped ions~\cite{brown2016co}. Moreover, the analysis highlights the importance of quantum repeater and router design. Since stochastic behavior normalizes at a larger scale, entanglement routing becomes predictable and readily buffered, improving efficiency and simplifying message passing. Multiplexed architecture with local optical switching is suitable for many memory types (e.g., using photonic circuits, spatial light modulators, and microelectromechanical mirrors). By combining low error correction overhead, scalability, and compatibility with classical telecom networks as well as near-term realistic hardware, the hierarchical routing and quantum tree network introduced here present a realistic blueprint to help guide quantum network architectures, protocols, and hardware development. 

\textit{Resilience against attacks and failures} - In a targeted attack, an attacker can disconnect the network by removing a small fraction of routers \cite{Magoni2003}. If only a portion of buffered qubits in a router failed, the network continued to operate with reduced performance. However, the tree network topology is vulnerable to router-wise attacks, errors, or other random failures. This is because each router is only connected to one node in the upper layer, so the removal of a router disconnects the end nodes connected to it from the network. 

A possible solution is to have $q$-redundant routers that form a Clos network with parent- and child-nodes. In the simplest case where every router is $q$-redundant, the total number of qubits increases $q^2$ times because each of $q$ routers has $q$ times the qubits to connect to $q$ times of upper and lower level routers. Meanwhile, the probability of failure decreases exponentially with $q$ if router failures are independent. Other hierarchical network structures, in addition to tree networks, can achieve sublinear scaling while offering resilience through redundant pathways.

\textit{Outlook} - Future works may consider other routing schemes, including Hamiltonian routing \cite{bapat2023advantages} or teleportation based routing \cite{devulapalli2022quantum}. In addition, one can incorporate ``quantum data center'' \cite{liu2022quantum} or quantum random access memory \cite{xu2023systems} to extend entanglement distribution to managing and processing a large scale quantum information. One can also improve QTN routing; further gains may be possible with mesh-tree hybrid graphs. 

\section*{acknowledgement}
We thank Yongshan Ding, Saikat Guha, Stephanie Wehner, Hassan Shapourian, Alireza Shabani, Stephen DiAdamo, Peter van Loock, Siddardha Chelluri, Don Towsley, Alexey Gorshkov, Junyu Liu, Liang Zhang, and Yufei Ding for fruitful discussions. This work was partially supported by AFOSR grant FA9550-20-1-0105 supervised by Gernot Pomrenke, the NSF Center for Quantum Networks (CQN, 1941583), and Cisco Research. H.C. and D.E. acknowledge HSBC, Mekena Metcalf, Zapata Computing, and Yudong Cao for MIT QSEC collaboration. H.C. and D.E. also acknowledge Honda Research Institute and Avetik Harutyunyan. H.C. acknowledges the Claude E. Shannon Fellowship and the Samsung Scholarship. M.G.D. acknowledges the DOE CSGF Fellowship.

H.C. and M.G.D contributed equally to this work.

\section*{Author Contribution}
D.E. initiated the idea of incompressible entanglement flow in a quantum network. D.E. and H.C. conceived the design of the quantum fat-tree network. H.C. studied the scalability of the quantum networks and derived the overheads. H.C. designed and M.G.D. performed the network simulation. H.C. and A.G.I. studied the network congestion and scaling of the network. A.G.I. studied the vulnerabilities and redundant protection. All authors provided critical feedback and helped shape the research, analysis, and manuscript.

\onecolumngrid

\begin{appendices}

\section{Congestion in mesh network}\label{apdx:CongestionMesh}

In this appendix, we investigate congestion in mesh networks. We consider a continuous plane, $(x,y) \in [0, 1]^2$, where the points represent nodes, and the straight lines defined by two endpoints are the routing paths. This setting generalizes any uniform lattice through the coarse-graining and the renormalization group. To represent uniform random traffic, we randomly choose pairs of points, and the straight line defined by a pair serves as the routing path of entanglement. Note that our following argument also applies to other mesh networks that are not based on uniform lattices, simply by adjusting the probability density of choosing points in the plane. 

If two routing paths intersect, the point of intersection needs at least a pair of quantum memories for entanglement swapping. Depending on the details of the intersection and the lattice, multiple additional pairs may be required. For example, if two lines intersect with a small angle in a coarse-grained space of a square lattice, the lines share multiple nodes in the original lattice. In the following, we consider the scaling of the number of intersections in the coarse-grained space, and the number of additional memories is lower-bounded by the number of intersections.

\begin{figure*}
    \centering
    \includegraphics[width=\textwidth]{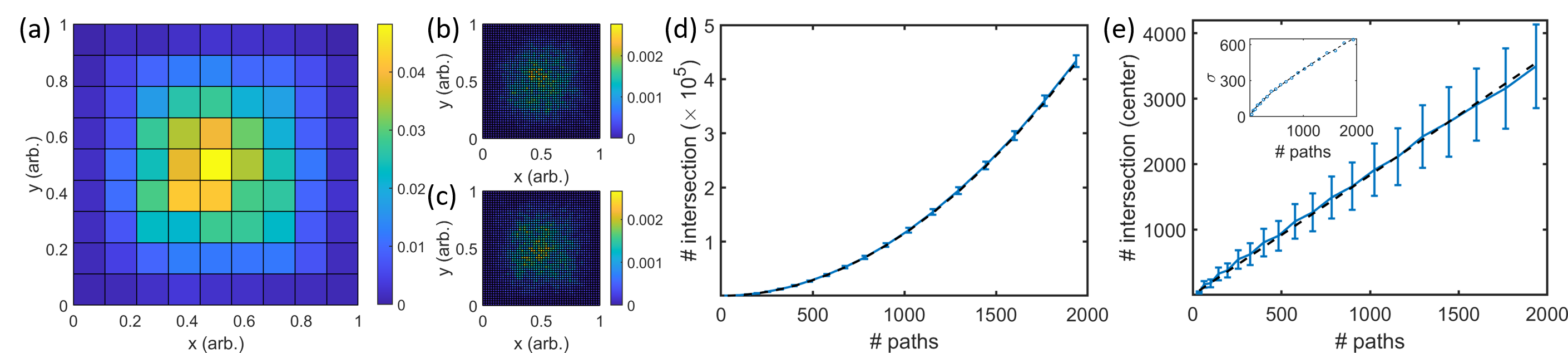}
    \caption{Congestion of routing paths in a continuous space. (a) Density map of intersections made by $N_e = 2,000$ routing paths. (b), (c) The instances of density of intersections in a finer $50\times 50$ grid. (d) The number of intersections plotted for $N_e$ paths. Each point is the average of 1,000 instances with plus-minus one standard-deviation error bars ($\pm\sigma$). Dashed line: $0.115 N_e^2$. (e) The number of intersections per node, at the center, is proportional to $N_e$ ($\approx 1.832N_e$). The inset shows the standard deviations, which grow as $2.109N_e^{0.757}$.}
    \label{fig:meshSim}
\end{figure*}

Figure~\ref{fig:meshSim}(a) shows the density map of the intersections made by the $N_e = 2,000$ lines. Each density in the $10\times 10$ grid is normalized by the total number of intersections in the space. The center area has a higher density of intersections. As we will confirm later, the area of high density is independent of $N_e$. In a mesh network, the density of intersections is not smooth in a finer grid (Fig.~\ref{fig:meshSim}(b), $50\times 50$ grid). The number of intersections is set to be much more than the number of grids ($4.3\times 10^5 \gg 50^2$), and the fluctuation of density across the grids is not caused by the discretization. There are fluctuations of densities across instances, as shown by comparing Fig.~\ref{fig:meshSim}(b) and (c). Because of the fluctuation, one needs to buffer each node with enough quantum memories to avoid congestion in the worst case scenario (see main text). Note that the worst-case for each node is still an element in the typical set of the set of routing paths over a mesh network.

Figure~\ref{fig:meshSim}(d) plots the number of intersections for different numbers of routing paths, $N_e$. Each data point is the average of 1,000 instances, and the error bars indicate one standard deviation of the samples ($\pm\sigma$). The black dashed line fits the data points with $0.115 N_e^2$. This confirms that the number of intersections in a mesh network follows as $\mathcal{O}(N_e^2)$, and $\sim\mathcal{O}(N^2)$ if $N_e \sim \mathcal{O}(N)$.

Additionally, we explored the number of intersections in the center with the area inversely proportional to $N_e$. Assuming $N_e \sim \mathcal{O}(N)$, the number of intersections within the area represents the number of intersections per node in the original lattice. Figure~\ref{fig:meshSim}(e) plots it at the center of $\sqrt{N_e}/2 \times \sqrt{N_e}/2$ grid with $\pm\sigma$ error bars. We fit the data points with the black dashed line, $y = 1.832N_e$ confirming that the intersection per node scales linearly with the number of routing paths. The inset plots the standard deviation of the number of intersections, which grows as $2.109N_e^{0.757}$.

\section{Activation probabilities}\label{apdx:ActProb}

Here, we calculate the activation probability of quantum routers in QTNs. The activation probability of a router is the probability that it needs to perform entanglement swapping for an entanglement request. Let us consider uniform random traffic where a randomly chosen one of $N$ end nodes requests entanglement with one of the others. This is equivalent to randomly choosing a pair out of all possible pairs. Table~\ref{tab:activationProb} shows the activation probability of a router labeled as N$x_0 ... x_y$. The probability that the router is on the routing path, but it is not the router at the highest layer of routing path, is $(1-p_{k,y}^2)\cdot p_{k,y}$ (``branch''), where $p_{k,y} = 1-\left(\frac{1}{k}\right)^y$ is the probability that two random $k$-ary numbers of length $y$ are different. On the other hand, the probability that a router is at the highest level of the routing path (``root'') is $k^{-2y}\cdot(k-1)/k$. The activation probability is the sum of these two contributions, and $\approx 2k^{-y}$ for a large $k$ or a large $y$ ($p_{k,y} \approx 1$). Thus, we can expect that the QTN will be congestion-free if the router has exponentially more qubits on the upper layers. Because the number of routers in the upper layer is exponentially smaller, the total number of qubits in each layer will be the same (Fig.~\ref{fig:TreeNetwork}(d)).

\begin{table}[]
    \centering
    \begin{tabular}{|c|c|c|c|c|}
     \hline
        router & kind & \multicolumn{2}{c|}{example pair} & probability* \\
     \hline
     \hline
        \multirow{2}{*}{N$x_0x_1...x_y$} & branch & \multirow{2}{*}{N$x_0x_1...x_y...x_n$} & N$x_0...x_i'...x_{y+1}''...x_n''$ & $(1-p^2_{k,y})\cdot p_{k,y}$ $\approx 2k^{-y}$\\ \cline{2-2} \cline{4-5}
                                         & root   &                                        & N$x_0x_1...x_yx_{y+1}'...x_n''$ & $\left(\frac{1}{k}\right)^{2y}\cdot\left(\frac{k-1}{k}\right)$\\
     \hline
    \end{tabular}
    \\ \vspace{1mm} *\footnotesize{two $k$-ary number of length $y$ to be different: $p_{k,y} = 1-\left(\frac{1}{k}\right)^y$, ``$\approx$'' for $p_{k,y} \approx 1$.}\\
    \caption{Activation Probability of $y+1^\text{th}$ layer router}
    \label{tab:activationProb}
\end{table}

\section{2D surface covering with Tree network}\label{apdx:2DCovering}

\begin{figure}
\centering
    \includegraphics[width=0.75\columnwidth]{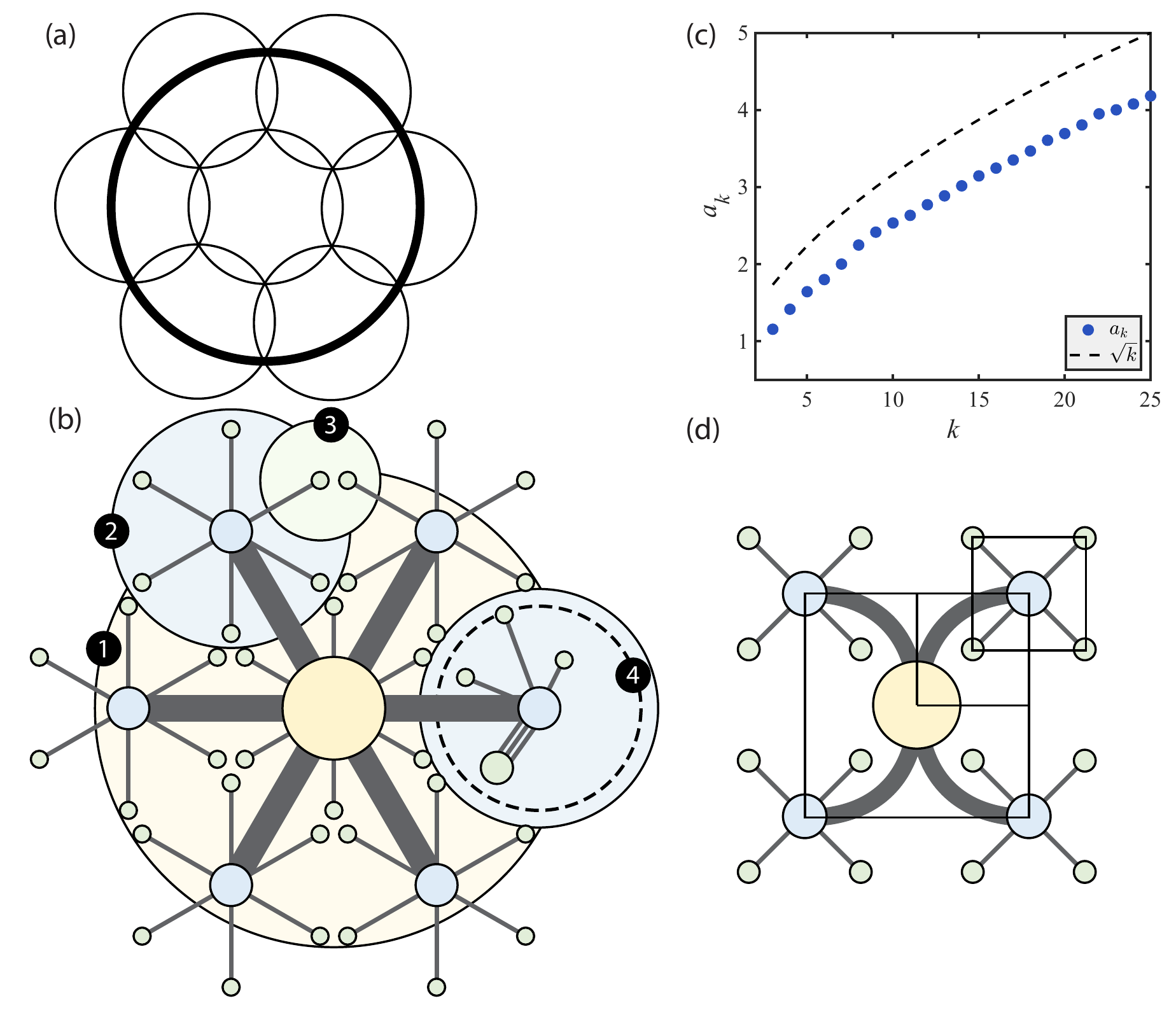}
    \caption{(a) Optimal solution to the disk covering problem for k=7.  (b) The layout of nodes in a QTN with $k=7$ and $n=2$, for a total of $N=49$.  Note that there is a blue, layer-1 node underneath the center layer-0 node, and there is a green end node under each of the blue, layer-1 nodes. (c) A plot of the values of $a_k$ for various values of $k$ based on the disk covering problem.  (d) A potential layout of a QTN with the end nodes organized in a grid layout. For this layout, $a_4^\text{sq}=2$, as demonstrated by the overlaid squares.}
    \label{fig:2D_covering}
\end{figure}

Quantum tree networks can fully cover the 2D surface to service the users in an area. Consider a $k$-ary uniform tree. We start from an end node having the area of service enclosed by a circle. The router nodes connected to an end node should cover the area dedicated to the router, also enclosed by a larger circle. The same process repeats for the higher layer's router, and the problem maps to the minimal disk covering problem~\cite{kershner1939number}. Figure~\ref{fig:2D_covering}(a) shows the $k=7$ disk covering problem. The disk covering problem asks for determining the maximum radius of a circle, whose area is covered by smaller $k$ circles of unit length. The solution to the minimal disk covering problems is analytically derived for a few $k$, and numerically known for the others~\cite{diskCovering}.

Figure~\ref{fig:2D_covering}(b) shows the application of the problem to QTNs. $k=7$ children nodes are spawned from the parent node across the tree. The green end nodes are the children of blue router nodes, and the blue router nodes are the children of yellow router nodes. Furthermore, the yellow nodes can be a child of an upper-layer router. The children are dispersed to cover the coverage of the parent node (\circled{1}). For every blue (yellow) node, one green (blue) node is not visible due to the overlap. The coverage assigned to the blue node (\circled{2}) is covered by the green nodes (\circled{3}). If network demands are highly concentrated in a local area, the nodes and the edges can be enhanced with multiplicities (\circled{4}). Note that the green nodes are not on the blue circle but are slightly inside the circle (\circled{4}). This is the same for the higher layer nodes, as the blue nodes are inside the yellow circle. The growth rate ($a_k$) is the ratio of the blue-green channel length to the green-yellow.

Figure~\ref{fig:2D_covering}(c) plots $a_k$ for different values of $k$s. As shown in the plot, $a_k$ slowly grows with $k$ and is always smaller than $\sqrt{k}$. The dashed line, $y=\sqrt{k}$, clearly shows this. One can easily prove that $a_k < \sqrt{k}$; the large circle cannot have a larger area than the sum of the areas of the $k$ smaller circles; $\pi (a_k)^2 < k\cdot \pi (1)^2$ $\rightarrow a_k < \sqrt{k}$. \emph{This guarantees the sub-square-root overhead of QTNs for any} $k$.

We call the aforementioned construction ``minimal surface covering'' in the main text, and used $a_k^\text{2D}$ for the growth rate. On the other hand, one can consider uniformly locating the end nodes at the positions of vertices on a square lattice. Figure \ref{fig:2D_covering}(d) shows the deployed nodes as well as the channels. The child nodes do not minimally cover the coverage of the parent. An end node covers the square area, within which the points have the node as the nearest one. The routers are symmetrically dispersed to connect the end nodes with a quaternary tree. We call this the ``square-lattice embedding'' construction, with $a_4^\text{sq} = 2$. Note that $a_4^\text{sq} \neq a_4^\text{2D} = \sqrt{2}$.

\section{Resource Overhead with Nested Error Correction}\label{apdx:resource_overhead}

For a $r$-nested error correction, each of which can correct up to $t$ errors, we can derive the logical error rate through induction;
\begin{align}
    \epsilon_L^{(r)} &= \epsilon_\text{th}\cdot \left(\frac{\epsilon}{\epsilon_\text{th}}\right)^{(t+1)^{r}}. \\
    t &\approx\sqrt[\mathlarger{\mathlarger{\mathlarger{\mathlarger{^r}}}}]{\frac{n\log a_k + \log(\epsilon/\epsilon_0)}{\log(\epsilon_\text{th}/\epsilon)}+1}-1. \label{eq:tForJ}
\end{align}
Equation~(\ref{eq:tForJ}) recovers Eq.~(\ref{eq:tForak}) for $r=1$. We can further simplify, assuming $n\log a_k \gg 1$. Moreover, we assume $\epsilon = 0.1\epsilon_\text{th}$ as in the main text. Then,
\begin{align}
    t &\approx \sqrt[r]{0.43n\log a_k}.
\end{align}
The total number of qubits on the network is,
\begin{align}
    \mathcal{N} = 2f(\mathcal{C},J,r,t)\cdot N \cdot (a_k^n-1)/(a_k-1),
\end{align}
where $f(\mathcal{C},J,r,t)$ is the encoding rate (physical qubits per logical qubit) and is a function of error correction code $\mathcal{C}$, nesting level $r$, and the code distance $2t+1$. Assuming a Gilbert-Varsharov bound-saturating code at each nesting level, 
\begin{align}
    \mathcal{N} \approx 0.86\frac{\log a_k}{a_k-1} \cdot 19^r \cdot N^{1+\log_k a_k} \cdot \log_k N. \label{eq:nested}
\end{align}
Note that it reconstructs Eq.~\ref{eq:overheadQTN} for $r=1$. This is in contrast to the result that the error correction overhead scales with $\left[\log(\text{distance})\right]^{r}$ shown in \cite{jiang2009quantum}. We attribute this discrepancy to our error normalization. We decided $t$ and changed $r$ for a fixed target error rate ($\epsilon_0$), while Jiang et. al. fixed $t$ and increased $r$ for a lower error rate \cite{jiang2009quantum}. There can be an alternative situation where physical constraints limit the maximum $t = t_\text{max}$, e.g. by a maximum weight of stabilizers, and a larger distance communication at a fixed error rate is only achieved by increasing $r$. In this case, 
\begin{align}
    r \approx \log\left(\frac{\log\text{(distance)}}{\log(\epsilon_\text{th}/\epsilon)}\right)\mathlarger{\mathlarger{\mathlarger{\mathlarger{/}}}}(t+1). \\
    \mathcal{N} \propto \text{distance}\cdot \left[\log\text{(distance)}\right]^{2.94} \label{eq:tJPhysConst}
\end{align}
where $2.94 = \log 19$ is from the encoding rate $n_\text{phys} \approx 19t$.

Table~\ref{tab:tJRelation} summarizes the comparison.

\begin{table}[h]
    \centering
    \begin{tabular}{|c|c|c|c|c|}
     \hline
        $t$ & $r$ & error normalization & scaling & note \\
     \hline
        variable & 1 & \checkmark & $\log L$ & Eq.~(\ref{eq:overheadQTN}) \\
     \hline
        variable & variable & \checkmark & $19^{r} \cdot \log L$ & Eq.~(\ref{eq:nested}) \\
     \hline
        fixed & variable & \xmark & $[\log L]^{r}$ & Ref.~\cite{jiang2009quantum} \\
    \hline
        $t_\text{max}$ & variable & \checkmark & $[\log L]^{2.94}$ & Eq.~(\ref{eq:tJPhysConst}) \\
    \hline
    \end{tabular}
    \caption{Cost coefficient scaling on distances ($L$: total distance)}
    \label{tab:tJRelation}
\end{table}

\section{Resource Overhead with Router-router encoding}\label{apdx:router-router_encoding}

In this section, we consider the scaling of the overhead with separate encoding of router-router channels. Here, we reverse the layer labeling and use $I = n - i \in {0,1,...,n}$, i.e. end nodes are in $I = 0^\text{th}$ layer. $I^\text{th}$-layer routers route entanglement from $k^I$ end nodes, and the raw entanglement generation rate required between the $I^\text{th}$ and $(I+1)^\text{th}$ layers is $\sim k^I \cdot R$ ($m=1$). The number of router stations at the $I^\text{th}$ layer is $N / k^I$, and the number of qubits in the repeater chain between the $I^\text{th}$ and $(I+1)^\text{th}$ layers is $\sim 2a_k^I \cdot k^I \cdot$ Poly$\left[\log(a_k^I)\right]$ (a factor of 2 comes from the two sides). Taking into account the total number of physical qubits in the interval $[I, I+1)$, we have $ \sim 2CN \cdot \left[\log(a_k)\right]^J \cdot a_k^I I^J = C'N a_k^I I^J$, where $J$ is the exponent of the encoding, $C$ is the constant (see main text and Appendix \ref{apdx:resource_overhead}), and $C'(a_k,J) = 2C\left[\log(a_k)\right]^J$. Taking into account the entire network, we find that the number of qubits is $\mathcal{N} \approx C'N \Sigma_{I=0}^{n-1}(a_k^I I^J) = C'N (\text{Li}_{-J}(a_k) - a_k^{n}\Phi(a_k, -J, n))$, where $\Phi$ denotes the Lerch transcendent, and $\text{Li}_s(z)$ is the Jonquiere function with complex order $s$, continued analytically to $z>1$. We can approximate and simplify transcendent functions; $\mathcal{N}< C'N(\Sigma_{I=0}^{n-1} a_k^I) \cdot (\Sigma_{I=0}^{n-1} I^J) < C''N \cdot a_k^n \cdot H_{n-1}^{(-J)} \approx C'' N^{1+\log_k(a_k)} \cdot (\log_kN)^{J+1}/(J+1)$, where $H_n^{(-J)}$ is the generalized harmonic number of order $-J$, and $C'' = C'/(a_k-1)$. As one can see, the overhead still scales sublinearly with the number of end nodes, but the exponent of logarithm is larger by one; $\mathcal{N}/N \sim \mathcal{O}\left(N^{\log_k a_k}\cdot(\log_k N)^{J+1}\right)$.

\section{Multipath finding}\label{apdx:multipath}

A central challenge in operating a networked quantum system with probabilistic links is that it requires the nodes to hold the memory until the end of the entanglement distribution to the end nodes. This raises the problem of efficient multipath finding within short communication and computation times. Finding and maximizing the number of independent paths on a graph with probabilistic edges, for a set of randomly generated requests, is challenging. Firstly, the computation requires global knowledge of the network, meaning that routers cannot make routing decisions without waiting for communication delays with other nodes. Secondly, the time complexity of finding paths grows super-linearly with the size of the graph. The specific time complexity depends on the characteristics of the graph itself and can vary. For more information on the time complexity of different graph types, refer to \cite{fredman1987fibonacci,thorup2000ram,pant2019routing}. 


\end{appendices}


\begin{thebibliography}{55}%
\makeatletter
\providecommand \@ifxundefined [1]{%
 \@ifx{#1\undefined}
}%
\providecommand \@ifnum [1]{%
 \ifnum #1\expandafter \@firstoftwo
 \else \expandafter \@secondoftwo
 \fi
}%
\providecommand \@ifx [1]{%
 \ifx #1\expandafter \@firstoftwo
 \else \expandafter \@secondoftwo
 \fi
}%
\providecommand \natexlab [1]{#1}%
\providecommand \enquote  [1]{``#1''}%
\providecommand \bibnamefont  [1]{#1}%
\providecommand \bibfnamefont [1]{#1}%
\providecommand \citenamefont [1]{#1}%
\providecommand \href@noop [0]{\@secondoftwo}%
\providecommand \href [0]{\begingroup \@sanitize@url \@href}%
\providecommand \@href[1]{\@@startlink{#1}\@@href}%
\providecommand \@@href[1]{\endgroup#1\@@endlink}%
\providecommand \@sanitize@url [0]{\catcode `\\12\catcode `\$12\catcode
  `\&12\catcode `\#12\catcode `\^12\catcode `\_12\catcode `\%12\relax}%
\providecommand \@@startlink[1]{}%
\providecommand \@@endlink[0]{}%
\providecommand \url  [0]{\begingroup\@sanitize@url \@url }%
\providecommand \@url [1]{\endgroup\@href {#1}{\urlprefix }}%
\providecommand \urlprefix  [0]{URL }%
\providecommand \Eprint [0]{\href }%
\providecommand \doibase [0]{https://doi.org/}%
\providecommand \selectlanguage [0]{\@gobble}%
\providecommand \bibinfo  [0]{\@secondoftwo}%
\providecommand \bibfield  [0]{\@secondoftwo}%
\providecommand \translation [1]{[#1]}%
\providecommand \BibitemOpen [0]{}%
\providecommand \bibitemStop [0]{}%
\providecommand \bibitemNoStop [0]{.\EOS\space}%
\providecommand \EOS [0]{\spacefactor3000\relax}%
\providecommand \BibitemShut  [1]{\csname bibitem#1\endcsname}%
\let\auto@bib@innerbib\@empty
\bibitem [{\citenamefont {Cirac}\ \emph {et~al.}(1997)\citenamefont {Cirac},
  \citenamefont {Zoller}, \citenamefont {Kimble},\ and\ \citenamefont
  {Mabuchi}}]{cirac1997quantum}%
  \BibitemOpen
  \bibfield  {author} {\bibinfo {author} {\bibfnamefont {J.~I.}\ \bibnamefont
  {Cirac}}, \bibinfo {author} {\bibfnamefont {P.}~\bibnamefont {Zoller}},
  \bibinfo {author} {\bibfnamefont {H.~J.}\ \bibnamefont {Kimble}},\ and\
  \bibinfo {author} {\bibfnamefont {H.}~\bibnamefont {Mabuchi}},\ }\href@noop
  {} {\bibfield  {journal} {\bibinfo  {journal} {Physical Review Letters}\
  }\textbf {\bibinfo {volume} {78}},\ \bibinfo {pages} {3221} (\bibinfo {year}
  {1997})}\BibitemShut {NoStop}%
\bibitem [{\citenamefont {Briegel}\ \emph {et~al.}(1998)\citenamefont
  {Briegel}, \citenamefont {D{\"u}r}, \citenamefont {Cirac},\ and\
  \citenamefont {Zoller}}]{briegel1998quantum}%
  \BibitemOpen
  \bibfield  {author} {\bibinfo {author} {\bibfnamefont {H.-J.}\ \bibnamefont
  {Briegel}}, \bibinfo {author} {\bibfnamefont {W.}~\bibnamefont {D{\"u}r}},
  \bibinfo {author} {\bibfnamefont {J.~I.}\ \bibnamefont {Cirac}},\ and\
  \bibinfo {author} {\bibfnamefont {P.}~\bibnamefont {Zoller}},\ }\href@noop {}
  {\bibfield  {journal} {\bibinfo  {journal} {Physical Review Letters}\
  }\textbf {\bibinfo {volume} {81}},\ \bibinfo {pages} {5932} (\bibinfo {year}
  {1998})}\BibitemShut {NoStop}%
\bibitem [{\citenamefont {Wehner}\ \emph {et~al.}(2018)\citenamefont {Wehner},
  \citenamefont {Elkouss},\ and\ \citenamefont {Hanson}}]{wehner2018quantum}%
  \BibitemOpen
  \bibfield  {author} {\bibinfo {author} {\bibfnamefont {S.}~\bibnamefont
  {Wehner}}, \bibinfo {author} {\bibfnamefont {D.}~\bibnamefont {Elkouss}},\
  and\ \bibinfo {author} {\bibfnamefont {R.}~\bibnamefont {Hanson}},\
  }\href@noop {} {\bibfield  {journal} {\bibinfo  {journal} {Science}\ }\textbf
  {\bibinfo {volume} {362}},\ \bibinfo {pages} {eaam9288} (\bibinfo {year}
  {2018})}\BibitemShut {NoStop}%
\bibitem [{\citenamefont {Ruf}\ \emph {et~al.}(2021)\citenamefont {Ruf},
  \citenamefont {Wan}, \citenamefont {Choi}, \citenamefont {Englund},\ and\
  \citenamefont {Hanson}}]{ruf2021quantum}%
  \BibitemOpen
  \bibfield  {author} {\bibinfo {author} {\bibfnamefont {M.}~\bibnamefont
  {Ruf}}, \bibinfo {author} {\bibfnamefont {N.~H.}\ \bibnamefont {Wan}},
  \bibinfo {author} {\bibfnamefont {H.}~\bibnamefont {Choi}}, \bibinfo {author}
  {\bibfnamefont {D.}~\bibnamefont {Englund}},\ and\ \bibinfo {author}
  {\bibfnamefont {R.}~\bibnamefont {Hanson}},\ }\href@noop {} {\bibfield
  {journal} {\bibinfo  {journal} {Journal of Applied Physics}\ }\textbf
  {\bibinfo {volume} {130}},\ \bibinfo {pages} {070901} (\bibinfo {year}
  {2021})}\BibitemShut {NoStop}%
\bibitem [{\citenamefont {Arrighi}\ and\ \citenamefont
  {Salvail}(2006)}]{arrighi2006blind}%
  \BibitemOpen
  \bibfield  {author} {\bibinfo {author} {\bibfnamefont {P.}~\bibnamefont
  {Arrighi}}\ and\ \bibinfo {author} {\bibfnamefont {L.}~\bibnamefont
  {Salvail}},\ }\href@noop {} {\bibfield  {journal} {\bibinfo  {journal}
  {International Journal of Quantum Information}\ }\textbf {\bibinfo {volume}
  {4}},\ \bibinfo {pages} {883} (\bibinfo {year} {2006})}\BibitemShut {NoStop}%
\bibitem [{\citenamefont {Zhang}\ and\ \citenamefont
  {Zhuang}(2021)}]{Zhang2021Distributed}%
  \BibitemOpen
  \bibfield  {author} {\bibinfo {author} {\bibfnamefont {Z.}~\bibnamefont
  {Zhang}}\ and\ \bibinfo {author} {\bibfnamefont {Q.}~\bibnamefont {Zhuang}},\
  }\href {https://doi.org/10.1088/2058-9565/abd4c3} {\bibfield  {journal}
  {\bibinfo  {journal} {Quantum Science and Technology}\ }\textbf {\bibinfo
  {volume} {6}},\ \bibinfo {pages} {043001} (\bibinfo {year}
  {2021})}\BibitemShut {NoStop}%
\bibitem [{\citenamefont {Hahn}\ \emph {et~al.}(2019)\citenamefont {Hahn},
  \citenamefont {Pappa},\ and\ \citenamefont {Eisert}}]{hahn2019quantum}%
  \BibitemOpen
  \bibfield  {author} {\bibinfo {author} {\bibfnamefont {F.}~\bibnamefont
  {Hahn}}, \bibinfo {author} {\bibfnamefont {A.}~\bibnamefont {Pappa}},\ and\
  \bibinfo {author} {\bibfnamefont {J.}~\bibnamefont {Eisert}},\ }\href@noop {}
  {\bibfield  {journal} {\bibinfo  {journal} {npj Quantum Information}\
  }\textbf {\bibinfo {volume} {5}},\ \bibinfo {pages} {1} (\bibinfo {year}
  {2019})}\BibitemShut {NoStop}%
\bibitem [{\citenamefont {Hayashi}\ \emph {et~al.}(2007)\citenamefont
  {Hayashi}, \citenamefont {Iwama}, \citenamefont {Nishimura}, \citenamefont
  {Raymond},\ and\ \citenamefont {Yamashita}}]{hayashi2007quantum}%
  \BibitemOpen
  \bibfield  {author} {\bibinfo {author} {\bibfnamefont {M.}~\bibnamefont
  {Hayashi}}, \bibinfo {author} {\bibfnamefont {K.}~\bibnamefont {Iwama}},
  \bibinfo {author} {\bibfnamefont {H.}~\bibnamefont {Nishimura}}, \bibinfo
  {author} {\bibfnamefont {R.}~\bibnamefont {Raymond}},\ and\ \bibinfo {author}
  {\bibfnamefont {S.}~\bibnamefont {Yamashita}},\ }in\ \href@noop {} {\emph
  {\bibinfo {booktitle} {STACS 2007: 24th Annual Symposium on Theoretical
  Aspects of Computer Science, Aachen, Germany, February 22-24, 2007.
  Proceedings 24}}}\ (\bibinfo {organization} {Springer},\ \bibinfo {year}
  {2007})\ pp.\ \bibinfo {pages} {610--621}\BibitemShut {NoStop}%
\bibitem [{\citenamefont {Pant}\ \emph {et~al.}(2019)\citenamefont {Pant},
  \citenamefont {Krovi}, \citenamefont {Towsley}, \citenamefont {Tassiulas},
  \citenamefont {Jiang}, \citenamefont {Basu}, \citenamefont {Englund},\ and\
  \citenamefont {Guha}}]{pant2019routing}%
  \BibitemOpen
  \bibfield  {author} {\bibinfo {author} {\bibfnamefont {M.}~\bibnamefont
  {Pant}}, \bibinfo {author} {\bibfnamefont {H.}~\bibnamefont {Krovi}},
  \bibinfo {author} {\bibfnamefont {D.}~\bibnamefont {Towsley}}, \bibinfo
  {author} {\bibfnamefont {L.}~\bibnamefont {Tassiulas}}, \bibinfo {author}
  {\bibfnamefont {L.}~\bibnamefont {Jiang}}, \bibinfo {author} {\bibfnamefont
  {P.}~\bibnamefont {Basu}}, \bibinfo {author} {\bibfnamefont {D.}~\bibnamefont
  {Englund}},\ and\ \bibinfo {author} {\bibfnamefont {S.}~\bibnamefont
  {Guha}},\ }\href@noop {} {\bibfield  {journal} {\bibinfo  {journal} {npj
  Quantum Information}\ }\textbf {\bibinfo {volume} {5}},\ \bibinfo {pages} {1}
  (\bibinfo {year} {2019})}\BibitemShut {NoStop}%
\bibitem [{\citenamefont {Patil}\ \emph {et~al.}(2022)\citenamefont {Patil},
  \citenamefont {Pant}, \citenamefont {Englund}, \citenamefont {Towsley},\ and\
  \citenamefont {Guha}}]{patil2022entanglement}%
  \BibitemOpen
  \bibfield  {author} {\bibinfo {author} {\bibfnamefont {A.}~\bibnamefont
  {Patil}}, \bibinfo {author} {\bibfnamefont {M.}~\bibnamefont {Pant}},
  \bibinfo {author} {\bibfnamefont {D.}~\bibnamefont {Englund}}, \bibinfo
  {author} {\bibfnamefont {D.}~\bibnamefont {Towsley}},\ and\ \bibinfo {author}
  {\bibfnamefont {S.}~\bibnamefont {Guha}},\ }\href@noop {} {\bibfield
  {journal} {\bibinfo  {journal} {npj Quantum Information}\ }\textbf {\bibinfo
  {volume} {8}},\ \bibinfo {pages} {1} (\bibinfo {year} {2022})}\BibitemShut
  {NoStop}%
\bibitem [{\citenamefont {Liu}\ \emph {et~al.}(2022)\citenamefont {Liu},
  \citenamefont {Hann},\ and\ \citenamefont {Jiang}}]{liu2022quantum}%
  \BibitemOpen
  \bibfield  {author} {\bibinfo {author} {\bibfnamefont {J.}~\bibnamefont
  {Liu}}, \bibinfo {author} {\bibfnamefont {C.~T.}\ \bibnamefont {Hann}},\ and\
  \bibinfo {author} {\bibfnamefont {L.}~\bibnamefont {Jiang}},\ }\href@noop {}
  {\bibfield  {journal} {\bibinfo  {journal} {arXiv preprint arXiv:2207.14336}\
  } (\bibinfo {year} {2022})}\BibitemShut {NoStop}%
\bibitem [{\citenamefont {Hasegawa}\ \emph {et~al.}(2018)\citenamefont
  {Hasegawa}, \citenamefont {Tamura}, \citenamefont {Sakuma}, \citenamefont
  {Kawaguchi}, \citenamefont {Yamamoto},\ and\ \citenamefont
  {Koyano}}]{hasegawa2018first}%
  \BibitemOpen
  \bibfield  {author} {\bibinfo {author} {\bibfnamefont {T.}~\bibnamefont
  {Hasegawa}}, \bibinfo {author} {\bibfnamefont {Y.}~\bibnamefont {Tamura}},
  \bibinfo {author} {\bibfnamefont {H.}~\bibnamefont {Sakuma}}, \bibinfo
  {author} {\bibfnamefont {Y.}~\bibnamefont {Kawaguchi}}, \bibinfo {author}
  {\bibfnamefont {Y.}~\bibnamefont {Yamamoto}},\ and\ \bibinfo {author}
  {\bibfnamefont {Y.}~\bibnamefont {Koyano}},\ }\href@noop {} {\bibfield
  {journal} {\bibinfo  {journal} {SEI Tech. Rev}\ }\textbf {\bibinfo {volume}
  {86}},\ \bibinfo {pages} {18} (\bibinfo {year} {2018})}\BibitemShut {NoStop}%
\bibitem [{\citenamefont {Rudolph}(2017)}]{rudolph2017optimistic}%
  \BibitemOpen
  \bibfield  {author} {\bibinfo {author} {\bibfnamefont {T.}~\bibnamefont
  {Rudolph}},\ }\href@noop {} {\bibfield  {journal} {\bibinfo  {journal} {APL
  photonics}\ }\textbf {\bibinfo {volume} {2}},\ \bibinfo {pages} {030901}
  (\bibinfo {year} {2017})}\BibitemShut {NoStop}%
\bibitem [{\citenamefont {Duan}\ \emph {et~al.}(2001)\citenamefont {Duan},
  \citenamefont {Lukin}, \citenamefont {Cirac},\ and\ \citenamefont
  {Zoller}}]{duan2001long}%
  \BibitemOpen
  \bibfield  {author} {\bibinfo {author} {\bibfnamefont {L.-M.}\ \bibnamefont
  {Duan}}, \bibinfo {author} {\bibfnamefont {M.~D.}\ \bibnamefont {Lukin}},
  \bibinfo {author} {\bibfnamefont {J.~I.}\ \bibnamefont {Cirac}},\ and\
  \bibinfo {author} {\bibfnamefont {P.}~\bibnamefont {Zoller}},\ }\href@noop {}
  {\bibfield  {journal} {\bibinfo  {journal} {Nature}\ }\textbf {\bibinfo
  {volume} {414}},\ \bibinfo {pages} {413} (\bibinfo {year}
  {2001})}\BibitemShut {NoStop}%
\bibitem [{\citenamefont {Barrett}\ and\ \citenamefont
  {Kok}(2005)}]{barrett2005efficient}%
  \BibitemOpen
  \bibfield  {author} {\bibinfo {author} {\bibfnamefont {S.~D.}\ \bibnamefont
  {Barrett}}\ and\ \bibinfo {author} {\bibfnamefont {P.}~\bibnamefont {Kok}},\
  }\href@noop {} {\bibfield  {journal} {\bibinfo  {journal} {Physical Review
  A}\ }\textbf {\bibinfo {volume} {71}},\ \bibinfo {pages} {060310} (\bibinfo
  {year} {2005})}\BibitemShut {NoStop}%
\bibitem [{\citenamefont {Bernien}\ \emph {et~al.}(2013)\citenamefont
  {Bernien}, \citenamefont {Hensen}, \citenamefont {Pfaff}, \citenamefont
  {Koolstra}, \citenamefont {Blok}, \citenamefont {Robledo}, \citenamefont
  {Taminiau}, \citenamefont {Markham}, \citenamefont {Twitchen}, \citenamefont
  {Childress} \emph {et~al.}}]{bernien2013heralded}%
  \BibitemOpen
  \bibfield  {author} {\bibinfo {author} {\bibfnamefont {H.}~\bibnamefont
  {Bernien}}, \bibinfo {author} {\bibfnamefont {B.}~\bibnamefont {Hensen}},
  \bibinfo {author} {\bibfnamefont {W.}~\bibnamefont {Pfaff}}, \bibinfo
  {author} {\bibfnamefont {G.}~\bibnamefont {Koolstra}}, \bibinfo {author}
  {\bibfnamefont {M.~S.}\ \bibnamefont {Blok}}, \bibinfo {author}
  {\bibfnamefont {L.}~\bibnamefont {Robledo}}, \bibinfo {author} {\bibfnamefont
  {T.~H.}\ \bibnamefont {Taminiau}}, \bibinfo {author} {\bibfnamefont
  {M.}~\bibnamefont {Markham}}, \bibinfo {author} {\bibfnamefont {D.~J.}\
  \bibnamefont {Twitchen}}, \bibinfo {author} {\bibfnamefont {L.}~\bibnamefont
  {Childress}}, \emph {et~al.},\ }\href@noop {} {\bibfield  {journal} {\bibinfo
   {journal} {Nature}\ }\textbf {\bibinfo {volume} {497}},\ \bibinfo {pages}
  {86} (\bibinfo {year} {2013})}\BibitemShut {NoStop}%
\bibitem [{\citenamefont {Bennett}\ \emph {et~al.}(1996)\citenamefont
  {Bennett}, \citenamefont {Brassard}, \citenamefont {Popescu}, \citenamefont
  {Schumacher}, \citenamefont {Smolin},\ and\ \citenamefont
  {Wootters}}]{bennett1996purification}%
  \BibitemOpen
  \bibfield  {author} {\bibinfo {author} {\bibfnamefont {C.~H.}\ \bibnamefont
  {Bennett}}, \bibinfo {author} {\bibfnamefont {G.}~\bibnamefont {Brassard}},
  \bibinfo {author} {\bibfnamefont {S.}~\bibnamefont {Popescu}}, \bibinfo
  {author} {\bibfnamefont {B.}~\bibnamefont {Schumacher}}, \bibinfo {author}
  {\bibfnamefont {J.~A.}\ \bibnamefont {Smolin}},\ and\ \bibinfo {author}
  {\bibfnamefont {W.~K.}\ \bibnamefont {Wootters}},\ }\href@noop {} {\bibfield
  {journal} {\bibinfo  {journal} {Physical review letters}\ }\textbf {\bibinfo
  {volume} {76}},\ \bibinfo {pages} {722} (\bibinfo {year} {1996})}\BibitemShut
  {NoStop}%
\bibitem [{\citenamefont {Jiang}\ \emph {et~al.}(2009)\citenamefont {Jiang},
  \citenamefont {Taylor}, \citenamefont {Nemoto}, \citenamefont {Munro},
  \citenamefont {Van~Meter},\ and\ \citenamefont {Lukin}}]{jiang2009quantum}%
  \BibitemOpen
  \bibfield  {author} {\bibinfo {author} {\bibfnamefont {L.}~\bibnamefont
  {Jiang}}, \bibinfo {author} {\bibfnamefont {J.~M.}\ \bibnamefont {Taylor}},
  \bibinfo {author} {\bibfnamefont {K.}~\bibnamefont {Nemoto}}, \bibinfo
  {author} {\bibfnamefont {W.~J.}\ \bibnamefont {Munro}}, \bibinfo {author}
  {\bibfnamefont {R.}~\bibnamefont {Van~Meter}},\ and\ \bibinfo {author}
  {\bibfnamefont {M.~D.}\ \bibnamefont {Lukin}},\ }\href@noop {} {\bibfield
  {journal} {\bibinfo  {journal} {Physical Review A}\ }\textbf {\bibinfo
  {volume} {79}},\ \bibinfo {pages} {032325} (\bibinfo {year}
  {2009})}\BibitemShut {NoStop}%
\bibitem [{\citenamefont {Munro}\ \emph {et~al.}(2012)\citenamefont {Munro},
  \citenamefont {Stephens}, \citenamefont {Devitt}, \citenamefont {Harrison},\
  and\ \citenamefont {Nemoto}}]{munro2012quantum}%
  \BibitemOpen
  \bibfield  {author} {\bibinfo {author} {\bibfnamefont {W.~J.}\ \bibnamefont
  {Munro}}, \bibinfo {author} {\bibfnamefont {A.~M.}\ \bibnamefont {Stephens}},
  \bibinfo {author} {\bibfnamefont {S.~J.}\ \bibnamefont {Devitt}}, \bibinfo
  {author} {\bibfnamefont {K.~A.}\ \bibnamefont {Harrison}},\ and\ \bibinfo
  {author} {\bibfnamefont {K.}~\bibnamefont {Nemoto}},\ }\href@noop {}
  {\bibfield  {journal} {\bibinfo  {journal} {Nature Photonics}\ }\textbf
  {\bibinfo {volume} {6}},\ \bibinfo {pages} {777} (\bibinfo {year}
  {2012})}\BibitemShut {NoStop}%
\bibitem [{\citenamefont {Roberts}(1978)}]{roberts1978evolution}%
  \BibitemOpen
  \bibfield  {author} {\bibinfo {author} {\bibfnamefont {L.~G.}\ \bibnamefont
  {Roberts}},\ }\href@noop {} {\bibfield  {journal} {\bibinfo  {journal}
  {Proceedings of the IEEE}\ }\textbf {\bibinfo {volume} {66}},\ \bibinfo
  {pages} {1307} (\bibinfo {year} {1978})}\BibitemShut {NoStop}%
\bibitem [{\citenamefont {Pompili}\ \emph {et~al.}(2021)\citenamefont
  {Pompili}, \citenamefont {Hermans}, \citenamefont {Baier}, \citenamefont
  {Beukers}, \citenamefont {Humphreys}, \citenamefont {Schouten}, \citenamefont
  {Vermeulen}, \citenamefont {Tiggelman}, \citenamefont {dos Santos~Martins},
  \citenamefont {Dirkse}, \citenamefont {Wehner},\ and\ \citenamefont
  {Hanson}}]{pompli2021}%
  \BibitemOpen
  \bibfield  {author} {\bibinfo {author} {\bibfnamefont {M.}~\bibnamefont
  {Pompili}}, \bibinfo {author} {\bibfnamefont {S.~L.~N.}\ \bibnamefont
  {Hermans}}, \bibinfo {author} {\bibfnamefont {S.}~\bibnamefont {Baier}},
  \bibinfo {author} {\bibfnamefont {H.~K.~C.}\ \bibnamefont {Beukers}},
  \bibinfo {author} {\bibfnamefont {P.~C.}\ \bibnamefont {Humphreys}}, \bibinfo
  {author} {\bibfnamefont {R.~N.}\ \bibnamefont {Schouten}}, \bibinfo {author}
  {\bibfnamefont {R.~F.~L.}\ \bibnamefont {Vermeulen}}, \bibinfo {author}
  {\bibfnamefont {M.~J.}\ \bibnamefont {Tiggelman}}, \bibinfo {author}
  {\bibfnamefont {L.}~\bibnamefont {dos Santos~Martins}}, \bibinfo {author}
  {\bibfnamefont {B.}~\bibnamefont {Dirkse}}, \bibinfo {author} {\bibfnamefont
  {S.}~\bibnamefont {Wehner}},\ and\ \bibinfo {author} {\bibfnamefont
  {R.}~\bibnamefont {Hanson}},\ }\href
  {https://doi.org/10.1126/science.abg1919} {\bibfield  {journal} {\bibinfo
  {journal} {Science}\ }\textbf {\bibinfo {volume} {372}},\ \bibinfo {pages}
  {259} (\bibinfo {year} {2021})}\BibitemShut {NoStop}%
\bibitem [{\citenamefont {Chow}\ \emph {et~al.}(2021)\citenamefont {Chow},
  \citenamefont {Dial},\ and\ \citenamefont {Gambetta}}]{chow2021ibm}%
  \BibitemOpen
  \bibfield  {author} {\bibinfo {author} {\bibfnamefont {J.}~\bibnamefont
  {Chow}}, \bibinfo {author} {\bibfnamefont {O.}~\bibnamefont {Dial}},\ and\
  \bibinfo {author} {\bibfnamefont {J.}~\bibnamefont {Gambetta}},\ }\href@noop
  {} {\bibfield  {journal} {\bibinfo  {journal} {IBM Research Blog}\ }
  (\bibinfo {year} {2021})}\BibitemShut {NoStop}%
\bibitem [{\citenamefont {Zahidy}\ \emph {et~al.}(2023)\citenamefont {Zahidy},
  \citenamefont {Mikkelsen}, \citenamefont {M{\"u}ller}, \citenamefont
  {Da~Lio}, \citenamefont {Krehbiel}, \citenamefont {Wang}, \citenamefont
  {Galili}, \citenamefont {Forchhammer}, \citenamefont {Lodahl}, \citenamefont
  {Oxenl{\o}we} \emph {et~al.}}]{zahidy2023quantum}%
  \BibitemOpen
  \bibfield  {author} {\bibinfo {author} {\bibfnamefont {M.}~\bibnamefont
  {Zahidy}}, \bibinfo {author} {\bibfnamefont {M.~T.}\ \bibnamefont
  {Mikkelsen}}, \bibinfo {author} {\bibfnamefont {R.}~\bibnamefont
  {M{\"u}ller}}, \bibinfo {author} {\bibfnamefont {B.}~\bibnamefont {Da~Lio}},
  \bibinfo {author} {\bibfnamefont {M.}~\bibnamefont {Krehbiel}}, \bibinfo
  {author} {\bibfnamefont {Y.}~\bibnamefont {Wang}}, \bibinfo {author}
  {\bibfnamefont {M.}~\bibnamefont {Galili}}, \bibinfo {author} {\bibfnamefont
  {S.}~\bibnamefont {Forchhammer}}, \bibinfo {author} {\bibfnamefont
  {P.}~\bibnamefont {Lodahl}}, \bibinfo {author} {\bibfnamefont {L.~K.}\
  \bibnamefont {Oxenl{\o}we}}, \emph {et~al.},\ }\href@noop {} {\bibfield
  {journal} {\bibinfo  {journal} {arXiv preprint arXiv:2301.09399}\ } (\bibinfo
  {year} {2023})}\BibitemShut {NoStop}%
\bibitem [{\citenamefont {Chakraborty}\ \emph {et~al.}(2020)\citenamefont
  {Chakraborty}, \citenamefont {Elkouss}, \citenamefont {Rijsman},\ and\
  \citenamefont {Wehner}}]{chakraborty2020entanglement}%
  \BibitemOpen
  \bibfield  {author} {\bibinfo {author} {\bibfnamefont {K.}~\bibnamefont
  {Chakraborty}}, \bibinfo {author} {\bibfnamefont {D.}~\bibnamefont
  {Elkouss}}, \bibinfo {author} {\bibfnamefont {B.}~\bibnamefont {Rijsman}},\
  and\ \bibinfo {author} {\bibfnamefont {S.}~\bibnamefont {Wehner}},\
  }\href@noop {} {\bibfield  {journal} {\bibinfo  {journal} {IEEE Transactions
  on Quantum Engineering}\ }\textbf {\bibinfo {volume} {1}},\ \bibinfo {pages}
  {1} (\bibinfo {year} {2020})}\BibitemShut {NoStop}%
\bibitem [{\citenamefont {Muralidharan}\ \emph {et~al.}(2016)\citenamefont
  {Muralidharan}, \citenamefont {Li}, \citenamefont {Kim}, \citenamefont
  {L{\"u}tkenhaus}, \citenamefont {Lukin},\ and\ \citenamefont
  {Jiang}}]{muralidharan2016optimal}%
  \BibitemOpen
  \bibfield  {author} {\bibinfo {author} {\bibfnamefont {S.}~\bibnamefont
  {Muralidharan}}, \bibinfo {author} {\bibfnamefont {L.}~\bibnamefont {Li}},
  \bibinfo {author} {\bibfnamefont {J.}~\bibnamefont {Kim}}, \bibinfo {author}
  {\bibfnamefont {N.}~\bibnamefont {L{\"u}tkenhaus}}, \bibinfo {author}
  {\bibfnamefont {M.~D.}\ \bibnamefont {Lukin}},\ and\ \bibinfo {author}
  {\bibfnamefont {L.}~\bibnamefont {Jiang}},\ }\href@noop {} {\bibfield
  {journal} {\bibinfo  {journal} {Scientific reports}\ }\textbf {\bibinfo
  {volume} {6}},\ \bibinfo {pages} {20463} (\bibinfo {year}
  {2016})}\BibitemShut {NoStop}%
\bibitem [{\citenamefont {Vardoyan}\ \emph {et~al.}(2019)\citenamefont
  {Vardoyan}, \citenamefont {Guha}, \citenamefont {Nain},\ and\ \citenamefont
  {Towsley}}]{vardoyan2019stochastic}%
  \BibitemOpen
  \bibfield  {author} {\bibinfo {author} {\bibfnamefont {G.}~\bibnamefont
  {Vardoyan}}, \bibinfo {author} {\bibfnamefont {S.}~\bibnamefont {Guha}},
  \bibinfo {author} {\bibfnamefont {P.}~\bibnamefont {Nain}},\ and\ \bibinfo
  {author} {\bibfnamefont {D.}~\bibnamefont {Towsley}},\ }\href@noop {}
  {\bibfield  {journal} {\bibinfo  {journal} {ACM SIGMETRICS Performance
  Evaluation Review}\ }\textbf {\bibinfo {volume} {47}},\ \bibinfo {pages} {27}
  (\bibinfo {year} {2019})}\BibitemShut {NoStop}%
\bibitem [{\citenamefont {Choi}\ \emph {et~al.}(2019)\citenamefont {Choi},
  \citenamefont {Pant}, \citenamefont {Guha},\ and\ \citenamefont
  {Englund}}]{choi2019percolation}%
  \BibitemOpen
  \bibfield  {author} {\bibinfo {author} {\bibfnamefont {H.}~\bibnamefont
  {Choi}}, \bibinfo {author} {\bibfnamefont {M.}~\bibnamefont {Pant}}, \bibinfo
  {author} {\bibfnamefont {S.}~\bibnamefont {Guha}},\ and\ \bibinfo {author}
  {\bibfnamefont {D.}~\bibnamefont {Englund}},\ }\href@noop {} {\bibfield
  {journal} {\bibinfo  {journal} {npj Quantum Information}\ }\textbf {\bibinfo
  {volume} {5}},\ \bibinfo {pages} {104} (\bibinfo {year} {2019})}\BibitemShut
  {NoStop}%
\bibitem [{\citenamefont {Munro}\ \emph {et~al.}(2010)\citenamefont {Munro},
  \citenamefont {Harrison}, \citenamefont {Stephens}, \citenamefont {Devitt},\
  and\ \citenamefont {Nemoto}}]{munro2010quantum}%
  \BibitemOpen
  \bibfield  {author} {\bibinfo {author} {\bibfnamefont {W.}~\bibnamefont
  {Munro}}, \bibinfo {author} {\bibfnamefont {K.}~\bibnamefont {Harrison}},
  \bibinfo {author} {\bibfnamefont {A.}~\bibnamefont {Stephens}}, \bibinfo
  {author} {\bibfnamefont {S.}~\bibnamefont {Devitt}},\ and\ \bibinfo {author}
  {\bibfnamefont {K.}~\bibnamefont {Nemoto}},\ }\href@noop {} {\bibfield
  {journal} {\bibinfo  {journal} {Nature Photonics}\ }\textbf {\bibinfo
  {volume} {4}},\ \bibinfo {pages} {792} (\bibinfo {year} {2010})}\BibitemShut
  {NoStop}%
\bibitem [{\citenamefont {Leiserson}(1985)}]{leiserson1985fat}%
  \BibitemOpen
  \bibfield  {author} {\bibinfo {author} {\bibfnamefont {C.~E.}\ \bibnamefont
  {Leiserson}},\ }\href@noop {} {\bibfield  {journal} {\bibinfo  {journal}
  {IEEE transactions on Computers}\ }\textbf {\bibinfo {volume} {100}},\
  \bibinfo {pages} {892} (\bibinfo {year} {1985})}\BibitemShut {NoStop}%
\bibitem [{\citenamefont {Al-Fares}\ \emph {et~al.}(2008)\citenamefont
  {Al-Fares}, \citenamefont {Loukissas},\ and\ \citenamefont
  {Vahdat}}]{al2008scalable}%
  \BibitemOpen
  \bibfield  {author} {\bibinfo {author} {\bibfnamefont {M.}~\bibnamefont
  {Al-Fares}}, \bibinfo {author} {\bibfnamefont {A.}~\bibnamefont
  {Loukissas}},\ and\ \bibinfo {author} {\bibfnamefont {A.}~\bibnamefont
  {Vahdat}},\ }\href@noop {} {\bibfield  {journal} {\bibinfo  {journal} {ACM
  SIGCOMM computer communication review}\ }\textbf {\bibinfo {volume} {38}},\
  \bibinfo {pages} {63} (\bibinfo {year} {2008})}\BibitemShut {NoStop}%
\bibitem [{\citenamefont {Lee}\ \emph {et~al.}(2022)\citenamefont {Lee},
  \citenamefont {Bersin}, \citenamefont {Dahlberg}, \citenamefont {Wehner},\
  and\ \citenamefont {Englund}}]{lee2022quantum}%
  \BibitemOpen
  \bibfield  {author} {\bibinfo {author} {\bibfnamefont {Y.}~\bibnamefont
  {Lee}}, \bibinfo {author} {\bibfnamefont {E.}~\bibnamefont {Bersin}},
  \bibinfo {author} {\bibfnamefont {A.}~\bibnamefont {Dahlberg}}, \bibinfo
  {author} {\bibfnamefont {S.}~\bibnamefont {Wehner}},\ and\ \bibinfo {author}
  {\bibfnamefont {D.}~\bibnamefont {Englund}},\ }\href@noop {} {\bibfield
  {journal} {\bibinfo  {journal} {npj Quantum Information}\ }\textbf {\bibinfo
  {volume} {8}},\ \bibinfo {pages} {1} (\bibinfo {year} {2022})}\BibitemShut
  {NoStop}%
\bibitem [{\citenamefont {Benjamin}\ \emph {et~al.}(2006)\citenamefont
  {Benjamin}, \citenamefont {Browne}, \citenamefont {Fitzsimons},\ and\
  \citenamefont {Morton}}]{benjamin2006brokered}%
  \BibitemOpen
  \bibfield  {author} {\bibinfo {author} {\bibfnamefont {S.~C.}\ \bibnamefont
  {Benjamin}}, \bibinfo {author} {\bibfnamefont {D.~E.}\ \bibnamefont
  {Browne}}, \bibinfo {author} {\bibfnamefont {J.}~\bibnamefont {Fitzsimons}},\
  and\ \bibinfo {author} {\bibfnamefont {J.~J.}\ \bibnamefont {Morton}},\
  }\href@noop {} {\bibfield  {journal} {\bibinfo  {journal} {New Journal of
  Physics}\ }\textbf {\bibinfo {volume} {8}},\ \bibinfo {pages} {141} (\bibinfo
  {year} {2006})}\BibitemShut {NoStop}%
\bibitem [{\citenamefont {Monroe}\ \emph {et~al.}(2014)\citenamefont {Monroe},
  \citenamefont {Raussendorf}, \citenamefont {Ruthven}, \citenamefont {Brown},
  \citenamefont {Maunz}, \citenamefont {Duan},\ and\ \citenamefont
  {Kim}}]{monroe2014large}%
  \BibitemOpen
  \bibfield  {author} {\bibinfo {author} {\bibfnamefont {C.}~\bibnamefont
  {Monroe}}, \bibinfo {author} {\bibfnamefont {R.}~\bibnamefont {Raussendorf}},
  \bibinfo {author} {\bibfnamefont {A.}~\bibnamefont {Ruthven}}, \bibinfo
  {author} {\bibfnamefont {K.~R.}\ \bibnamefont {Brown}}, \bibinfo {author}
  {\bibfnamefont {P.}~\bibnamefont {Maunz}}, \bibinfo {author} {\bibfnamefont
  {L.-M.}\ \bibnamefont {Duan}},\ and\ \bibinfo {author} {\bibfnamefont
  {J.}~\bibnamefont {Kim}},\ }\href@noop {} {\bibfield  {journal} {\bibinfo
  {journal} {Physical Review A}\ }\textbf {\bibinfo {volume} {89}},\ \bibinfo
  {pages} {022317} (\bibinfo {year} {2014})}\BibitemShut {NoStop}%
\bibitem [{\citenamefont {Brown}\ \emph {et~al.}(2016)\citenamefont {Brown},
  \citenamefont {Kim},\ and\ \citenamefont {Monroe}}]{brown2016co}%
  \BibitemOpen
  \bibfield  {author} {\bibinfo {author} {\bibfnamefont {K.~R.}\ \bibnamefont
  {Brown}}, \bibinfo {author} {\bibfnamefont {J.}~\bibnamefont {Kim}},\ and\
  \bibinfo {author} {\bibfnamefont {C.}~\bibnamefont {Monroe}},\ }\href@noop {}
  {\bibfield  {journal} {\bibinfo  {journal} {npj Quantum Information}\
  }\textbf {\bibinfo {volume} {2}},\ \bibinfo {pages} {1} (\bibinfo {year}
  {2016})}\BibitemShut {NoStop}%
\bibitem [{\citenamefont {Wan}\ \emph {et~al.}(2020)\citenamefont {Wan},
  \citenamefont {Lu}, \citenamefont {Chen}, \citenamefont {Walsh},
  \citenamefont {Trusheim}, \citenamefont {De~Santis}, \citenamefont {Bersin},
  \citenamefont {Harris}, \citenamefont {Mouradian}, \citenamefont {Christen}
  \emph {et~al.}}]{wan2020large}%
  \BibitemOpen
  \bibfield  {author} {\bibinfo {author} {\bibfnamefont {N.~H.}\ \bibnamefont
  {Wan}}, \bibinfo {author} {\bibfnamefont {T.-J.}\ \bibnamefont {Lu}},
  \bibinfo {author} {\bibfnamefont {K.~C.}\ \bibnamefont {Chen}}, \bibinfo
  {author} {\bibfnamefont {M.~P.}\ \bibnamefont {Walsh}}, \bibinfo {author}
  {\bibfnamefont {M.~E.}\ \bibnamefont {Trusheim}}, \bibinfo {author}
  {\bibfnamefont {L.}~\bibnamefont {De~Santis}}, \bibinfo {author}
  {\bibfnamefont {E.~A.}\ \bibnamefont {Bersin}}, \bibinfo {author}
  {\bibfnamefont {I.~B.}\ \bibnamefont {Harris}}, \bibinfo {author}
  {\bibfnamefont {S.~L.}\ \bibnamefont {Mouradian}}, \bibinfo {author}
  {\bibfnamefont {I.~R.}\ \bibnamefont {Christen}}, \emph {et~al.},\
  }\href@noop {} {\bibfield  {journal} {\bibinfo  {journal} {Nature}\ }\textbf
  {\bibinfo {volume} {583}},\ \bibinfo {pages} {226} (\bibinfo {year}
  {2020})}\BibitemShut {NoStop}%
\bibitem [{\citenamefont {Li}\ \emph {et~al.}(2022)\citenamefont {Li},
  \citenamefont {De~Santis}, \citenamefont {Harris}, \citenamefont {Chen},
  \citenamefont {Song}, \citenamefont {Christen}, \citenamefont {Trusheim},
  \citenamefont {Herranz}, \citenamefont {Han},\ and\ \citenamefont
  {Englund}}]{li2022scalable}%
  \BibitemOpen
  \bibfield  {author} {\bibinfo {author} {\bibfnamefont {L.}~\bibnamefont
  {Li}}, \bibinfo {author} {\bibfnamefont {L.}~\bibnamefont {De~Santis}},
  \bibinfo {author} {\bibfnamefont {I.}~\bibnamefont {Harris}}, \bibinfo
  {author} {\bibfnamefont {K.}~\bibnamefont {Chen}}, \bibinfo {author}
  {\bibfnamefont {Y.}~\bibnamefont {Song}}, \bibinfo {author} {\bibfnamefont
  {I.}~\bibnamefont {Christen}}, \bibinfo {author} {\bibfnamefont
  {M.}~\bibnamefont {Trusheim}}, \bibinfo {author} {\bibfnamefont {C.~E.}\
  \bibnamefont {Herranz}}, \bibinfo {author} {\bibfnamefont {R.}~\bibnamefont
  {Han}},\ and\ \bibinfo {author} {\bibfnamefont {D.}~\bibnamefont {Englund}},\
  }in\ \href@noop {} {\emph {\bibinfo {booktitle} {2022 Conference on Lasers
  and Electro-Optics (CLEO)}}}\ (\bibinfo {organization} {IEEE},\ \bibinfo
  {year} {2022})\ pp.\ \bibinfo {pages} {1--2}\BibitemShut {NoStop}%
\bibitem [{\citenamefont {Calderbank}\ and\ \citenamefont
  {Shor}(1996)}]{calderbank1996good}%
  \BibitemOpen
  \bibfield  {author} {\bibinfo {author} {\bibfnamefont {A.~R.}\ \bibnamefont
  {Calderbank}}\ and\ \bibinfo {author} {\bibfnamefont {P.~W.}\ \bibnamefont
  {Shor}},\ }\href@noop {} {\bibfield  {journal} {\bibinfo  {journal} {Physical
  Review A}\ }\textbf {\bibinfo {volume} {54}},\ \bibinfo {pages} {1098}
  (\bibinfo {year} {1996})}\BibitemShut {NoStop}%
\bibitem [{\citenamefont {Nielsen}\ and\ \citenamefont
  {Chuang}(2002)}]{nielsen2002quantum}%
  \BibitemOpen
  \bibfield  {author} {\bibinfo {author} {\bibfnamefont {M.~A.}\ \bibnamefont
  {Nielsen}}\ and\ \bibinfo {author} {\bibfnamefont {I.}~\bibnamefont
  {Chuang}},\ }\href@noop {} {\bibinfo {title} {Quantum computation and quantum
  information}} (\bibinfo {year} {2002})\BibitemShut {NoStop}%
\bibitem [{\citenamefont {Fowler}\ \emph {et~al.}(2012)\citenamefont {Fowler},
  \citenamefont {Mariantoni}, \citenamefont {Martinis},\ and\ \citenamefont
  {Cleland}}]{fowler2012surface}%
  \BibitemOpen
  \bibfield  {author} {\bibinfo {author} {\bibfnamefont {A.~G.}\ \bibnamefont
  {Fowler}}, \bibinfo {author} {\bibfnamefont {M.}~\bibnamefont {Mariantoni}},
  \bibinfo {author} {\bibfnamefont {J.~M.}\ \bibnamefont {Martinis}},\ and\
  \bibinfo {author} {\bibfnamefont {A.~N.}\ \bibnamefont {Cleland}},\
  }\href@noop {} {\bibfield  {journal} {\bibinfo  {journal} {Physical Review
  A}\ }\textbf {\bibinfo {volume} {86}},\ \bibinfo {pages} {032324} (\bibinfo
  {year} {2012})}\BibitemShut {NoStop}%
\bibitem [{\citenamefont {Arute}\ \emph {et~al.}(2019)\citenamefont {Arute},
  \citenamefont {Arya}, \citenamefont {Babbush}, \citenamefont {Bacon},
  \citenamefont {Bardin}, \citenamefont {Barends}, \citenamefont {Biswas},
  \citenamefont {Boixo}, \citenamefont {Brandao}, \citenamefont {Buell} \emph
  {et~al.}}]{arute2019quantum}%
  \BibitemOpen
  \bibfield  {author} {\bibinfo {author} {\bibfnamefont {F.}~\bibnamefont
  {Arute}}, \bibinfo {author} {\bibfnamefont {K.}~\bibnamefont {Arya}},
  \bibinfo {author} {\bibfnamefont {R.}~\bibnamefont {Babbush}}, \bibinfo
  {author} {\bibfnamefont {D.}~\bibnamefont {Bacon}}, \bibinfo {author}
  {\bibfnamefont {J.~C.}\ \bibnamefont {Bardin}}, \bibinfo {author}
  {\bibfnamefont {R.}~\bibnamefont {Barends}}, \bibinfo {author} {\bibfnamefont
  {R.}~\bibnamefont {Biswas}}, \bibinfo {author} {\bibfnamefont
  {S.}~\bibnamefont {Boixo}}, \bibinfo {author} {\bibfnamefont {F.~G.}\
  \bibnamefont {Brandao}}, \bibinfo {author} {\bibfnamefont {D.~A.}\
  \bibnamefont {Buell}}, \emph {et~al.},\ }\href@noop {} {\bibfield  {journal}
  {\bibinfo  {journal} {Nature}\ }\textbf {\bibinfo {volume} {574}},\ \bibinfo
  {pages} {505} (\bibinfo {year} {2019})}\BibitemShut {NoStop}%
\bibitem [{\citenamefont {Bapat}\ \emph {et~al.}(2018)\citenamefont {Bapat},
  \citenamefont {Eldredge}, \citenamefont {Garrison}, \citenamefont
  {Deshpande}, \citenamefont {Chong},\ and\ \citenamefont
  {Gorshkov}}]{bapat2018unitary}%
  \BibitemOpen
  \bibfield  {author} {\bibinfo {author} {\bibfnamefont {A.}~\bibnamefont
  {Bapat}}, \bibinfo {author} {\bibfnamefont {Z.}~\bibnamefont {Eldredge}},
  \bibinfo {author} {\bibfnamefont {J.~R.}\ \bibnamefont {Garrison}}, \bibinfo
  {author} {\bibfnamefont {A.}~\bibnamefont {Deshpande}}, \bibinfo {author}
  {\bibfnamefont {F.~T.}\ \bibnamefont {Chong}},\ and\ \bibinfo {author}
  {\bibfnamefont {A.~V.}\ \bibnamefont {Gorshkov}},\ }\href@noop {} {\bibfield
  {journal} {\bibinfo  {journal} {Physical Review A}\ }\textbf {\bibinfo
  {volume} {98}},\ \bibinfo {pages} {062328} (\bibinfo {year}
  {2018})}\BibitemShut {NoStop}%
\bibitem [{\citenamefont {Eldredge}\ \emph {et~al.}(2020)\citenamefont
  {Eldredge}, \citenamefont {Zhou}, \citenamefont {Bapat}, \citenamefont
  {Garrison}, \citenamefont {Deshpande}, \citenamefont {Chong},\ and\
  \citenamefont {Gorshkov}}]{eldredge2020entanglement}%
  \BibitemOpen
  \bibfield  {author} {\bibinfo {author} {\bibfnamefont {Z.}~\bibnamefont
  {Eldredge}}, \bibinfo {author} {\bibfnamefont {L.}~\bibnamefont {Zhou}},
  \bibinfo {author} {\bibfnamefont {A.}~\bibnamefont {Bapat}}, \bibinfo
  {author} {\bibfnamefont {J.~R.}\ \bibnamefont {Garrison}}, \bibinfo {author}
  {\bibfnamefont {A.}~\bibnamefont {Deshpande}}, \bibinfo {author}
  {\bibfnamefont {F.~T.}\ \bibnamefont {Chong}},\ and\ \bibinfo {author}
  {\bibfnamefont {A.~V.}\ \bibnamefont {Gorshkov}},\ }\href@noop {} {\bibfield
  {journal} {\bibinfo  {journal} {Physical review research}\ }\textbf {\bibinfo
  {volume} {2}},\ \bibinfo {pages} {033316} (\bibinfo {year}
  {2020})}\BibitemShut {NoStop}%
\bibitem [{\citenamefont {Davis}\ \emph {et~al.}(2023)\citenamefont {Davis},
  \citenamefont {Choi}, \citenamefont {I{\~n}esta~G},\ and\ \citenamefont
  {Englund}}]{Davis2023QTN}%
  \BibitemOpen
  \bibfield  {author} {\bibinfo {author} {\bibfnamefont {M.}~\bibnamefont
  {Davis}}, \bibinfo {author} {\bibfnamefont {H.}~\bibnamefont {Choi}},
  \bibinfo {author} {\bibfnamefont {{\'A}.}~\bibnamefont {I{\~n}esta~G}},\ and\
  \bibinfo {author} {\bibfnamefont {D.}~\bibnamefont {Englund}},\ }\href@noop
  {} {\bibinfo {title} {quantum tree network}},\ \bibinfo {howpublished}
  {\url{https://github.com/WolfLink/quantum_tree_network_simulator}} (\bibinfo
  {year} {2023})\BibitemShut {NoStop}%
\bibitem [{\citenamefont {Humphreys}\ \emph {et~al.}(2018)\citenamefont
  {Humphreys}, \citenamefont {Kalb}, \citenamefont {Morits}, \citenamefont
  {Schouten}, \citenamefont {Vermeulen}, \citenamefont {Twitchen},
  \citenamefont {Markham},\ and\ \citenamefont
  {Hanson}}]{humphreys2018deterministic}%
  \BibitemOpen
  \bibfield  {author} {\bibinfo {author} {\bibfnamefont {P.~C.}\ \bibnamefont
  {Humphreys}}, \bibinfo {author} {\bibfnamefont {N.}~\bibnamefont {Kalb}},
  \bibinfo {author} {\bibfnamefont {J.~P.}\ \bibnamefont {Morits}}, \bibinfo
  {author} {\bibfnamefont {R.~N.}\ \bibnamefont {Schouten}}, \bibinfo {author}
  {\bibfnamefont {R.~F.}\ \bibnamefont {Vermeulen}}, \bibinfo {author}
  {\bibfnamefont {D.~J.}\ \bibnamefont {Twitchen}}, \bibinfo {author}
  {\bibfnamefont {M.}~\bibnamefont {Markham}},\ and\ \bibinfo {author}
  {\bibfnamefont {R.}~\bibnamefont {Hanson}},\ }\href@noop {} {\bibfield
  {journal} {\bibinfo  {journal} {Nature}\ }\textbf {\bibinfo {volume} {558}},\
  \bibinfo {pages} {268} (\bibinfo {year} {2018})}\BibitemShut {NoStop}%
\bibitem [{\citenamefont {Bradley}\ \emph {et~al.}(2019)\citenamefont
  {Bradley}, \citenamefont {Randall}, \citenamefont {Abobeih}, \citenamefont
  {Berrevoets}, \citenamefont {Degen}, \citenamefont {Bakker}, \citenamefont
  {Markham}, \citenamefont {Twitchen},\ and\ \citenamefont
  {Taminiau}}]{bradley2019ten}%
  \BibitemOpen
  \bibfield  {author} {\bibinfo {author} {\bibfnamefont {C.~E.}\ \bibnamefont
  {Bradley}}, \bibinfo {author} {\bibfnamefont {J.}~\bibnamefont {Randall}},
  \bibinfo {author} {\bibfnamefont {M.~H.}\ \bibnamefont {Abobeih}}, \bibinfo
  {author} {\bibfnamefont {R.}~\bibnamefont {Berrevoets}}, \bibinfo {author}
  {\bibfnamefont {M.}~\bibnamefont {Degen}}, \bibinfo {author} {\bibfnamefont
  {M.~A.}\ \bibnamefont {Bakker}}, \bibinfo {author} {\bibfnamefont
  {M.}~\bibnamefont {Markham}}, \bibinfo {author} {\bibfnamefont
  {D.}~\bibnamefont {Twitchen}},\ and\ \bibinfo {author} {\bibfnamefont
  {T.~H.}\ \bibnamefont {Taminiau}},\ }\href@noop {} {\bibfield  {journal}
  {\bibinfo  {journal} {Physical Review X}\ }\textbf {\bibinfo {volume} {9}},\
  \bibinfo {pages} {031045} (\bibinfo {year} {2019})}\BibitemShut {NoStop}%
\bibitem [{\citenamefont {Li}\ \emph {et~al.}(2023)\citenamefont {Li},
  \citenamefont {Li}, \citenamefont {Xue}, \citenamefont {Wei}, \citenamefont
  {Li}, \citenamefont {Yu}, \citenamefont {Sun},\ and\ \citenamefont
  {Lu}}]{li2023swapping}%
  \BibitemOpen
  \bibfield  {author} {\bibinfo {author} {\bibfnamefont {Z.}~\bibnamefont
  {Li}}, \bibinfo {author} {\bibfnamefont {J.}~\bibnamefont {Li}}, \bibinfo
  {author} {\bibfnamefont {K.}~\bibnamefont {Xue}}, \bibinfo {author}
  {\bibfnamefont {D.~S.}\ \bibnamefont {Wei}}, \bibinfo {author} {\bibfnamefont
  {R.}~\bibnamefont {Li}}, \bibinfo {author} {\bibfnamefont {N.}~\bibnamefont
  {Yu}}, \bibinfo {author} {\bibfnamefont {Q.}~\bibnamefont {Sun}},\ and\
  \bibinfo {author} {\bibfnamefont {J.}~\bibnamefont {Lu}},\ }\href@noop {}
  {\bibfield  {journal} {\bibinfo  {journal} {IEEE Transactions on Network and
  Service Management}\ } (\bibinfo {year} {2023})}\BibitemShut {NoStop}%
\bibitem [{\citenamefont {Chen}\ \emph {et~al.}(2023)\citenamefont {Chen},
  \citenamefont {Xue}, \citenamefont {Li}, \citenamefont {Li}, \citenamefont
  {Yu}, \citenamefont {Sun},\ and\ \citenamefont {Lu}}]{chen2023q}%
  \BibitemOpen
  \bibfield  {author} {\bibinfo {author} {\bibfnamefont {L.}~\bibnamefont
  {Chen}}, \bibinfo {author} {\bibfnamefont {K.}~\bibnamefont {Xue}}, \bibinfo
  {author} {\bibfnamefont {J.}~\bibnamefont {Li}}, \bibinfo {author}
  {\bibfnamefont {R.}~\bibnamefont {Li}}, \bibinfo {author} {\bibfnamefont
  {N.}~\bibnamefont {Yu}}, \bibinfo {author} {\bibfnamefont {Q.}~\bibnamefont
  {Sun}},\ and\ \bibinfo {author} {\bibfnamefont {J.}~\bibnamefont {Lu}},\
  }\href@noop {} {\bibfield  {journal} {\bibinfo  {journal} {IEEE/ACM
  Transactions on Networking}\ } (\bibinfo {year} {2023})}\BibitemShut
  {NoStop}%
\bibitem [{\citenamefont {Magoni}(2003)}]{Magoni2003}%
  \BibitemOpen
  \bibfield  {author} {\bibinfo {author} {\bibfnamefont {D.}~\bibnamefont
  {Magoni}},\ }\href@noop {} {\bibfield  {journal} {\bibinfo  {journal} {IEEE
  Journal on Selected Areas in Communications}\ }\textbf {\bibinfo {volume}
  {21}},\ \bibinfo {pages} {949} (\bibinfo {year} {2003})}\BibitemShut
  {NoStop}%
\bibitem [{\citenamefont {Bapat}\ \emph {et~al.}(2023)\citenamefont {Bapat},
  \citenamefont {Childs}, \citenamefont {Gorshkov},\ and\ \citenamefont
  {Schoute}}]{bapat2023advantages}%
  \BibitemOpen
  \bibfield  {author} {\bibinfo {author} {\bibfnamefont {A.}~\bibnamefont
  {Bapat}}, \bibinfo {author} {\bibfnamefont {A.~M.}\ \bibnamefont {Childs}},
  \bibinfo {author} {\bibfnamefont {A.~V.}\ \bibnamefont {Gorshkov}},\ and\
  \bibinfo {author} {\bibfnamefont {E.}~\bibnamefont {Schoute}},\ }\href@noop
  {} {\bibfield  {journal} {\bibinfo  {journal} {PRX Quantum}\ }\textbf
  {\bibinfo {volume} {4}},\ \bibinfo {pages} {010313} (\bibinfo {year}
  {2023})}\BibitemShut {NoStop}%
\bibitem [{\citenamefont {Devulapalli}\ \emph {et~al.}(2022)\citenamefont
  {Devulapalli}, \citenamefont {Schoute}, \citenamefont {Bapat}, \citenamefont
  {Childs},\ and\ \citenamefont {Gorshkov}}]{devulapalli2022quantum}%
  \BibitemOpen
  \bibfield  {author} {\bibinfo {author} {\bibfnamefont {D.}~\bibnamefont
  {Devulapalli}}, \bibinfo {author} {\bibfnamefont {E.}~\bibnamefont
  {Schoute}}, \bibinfo {author} {\bibfnamefont {A.}~\bibnamefont {Bapat}},
  \bibinfo {author} {\bibfnamefont {A.~M.}\ \bibnamefont {Childs}},\ and\
  \bibinfo {author} {\bibfnamefont {A.~V.}\ \bibnamefont {Gorshkov}},\
  }\href@noop {} {\bibfield  {journal} {\bibinfo  {journal} {arXiv preprint
  arXiv:2204.04185}\ } (\bibinfo {year} {2022})}\BibitemShut {NoStop}%
\bibitem [{\citenamefont {Xu}\ \emph {et~al.}(2023)\citenamefont {Xu},
  \citenamefont {Hann}, \citenamefont {Foxman}, \citenamefont {Girvin},\ and\
  \citenamefont {Ding}}]{xu2023systems}%
  \BibitemOpen
  \bibfield  {author} {\bibinfo {author} {\bibfnamefont {S.}~\bibnamefont
  {Xu}}, \bibinfo {author} {\bibfnamefont {C.~T.}\ \bibnamefont {Hann}},
  \bibinfo {author} {\bibfnamefont {B.}~\bibnamefont {Foxman}}, \bibinfo
  {author} {\bibfnamefont {S.~M.}\ \bibnamefont {Girvin}},\ and\ \bibinfo
  {author} {\bibfnamefont {Y.}~\bibnamefont {Ding}},\ }\href@noop {} {\bibfield
   {journal} {\bibinfo  {journal} {arXiv preprint arXiv:2306.03242}\ }
  (\bibinfo {year} {2023})}\BibitemShut {NoStop}%
\bibitem [{\citenamefont {Kershner}(1939)}]{kershner1939number}%
  \BibitemOpen
  \bibfield  {author} {\bibinfo {author} {\bibfnamefont {R.}~\bibnamefont
  {Kershner}},\ }\href@noop {} {\bibfield  {journal} {\bibinfo  {journal}
  {American Journal of mathematics}\ }\textbf {\bibinfo {volume} {61}},\
  \bibinfo {pages} {665} (\bibinfo {year} {1939})}\BibitemShut {NoStop}%
\bibitem [{\citenamefont {Friedman}()}]{diskCovering}%
  \BibitemOpen
  \bibfield  {author} {\bibinfo {author} {\bibfnamefont {E.}~\bibnamefont
  {Friedman}},\ }\href@noop {} {\bibinfo {title} {Circles covering circles}},\
  \bibinfo {howpublished}
  {\url{https://erich-friedman.github.io/packing/circovcir/}},\ \bibinfo {note}
  {accessed: 2023-06-10}\BibitemShut {NoStop}%
\bibitem [{\citenamefont {Fredman}\ and\ \citenamefont
  {Tarjan}(1987)}]{fredman1987fibonacci}%
  \BibitemOpen
  \bibfield  {author} {\bibinfo {author} {\bibfnamefont {M.~L.}\ \bibnamefont
  {Fredman}}\ and\ \bibinfo {author} {\bibfnamefont {R.~E.}\ \bibnamefont
  {Tarjan}},\ }\href@noop {} {\bibfield  {journal} {\bibinfo  {journal}
  {Journal of the ACM (JACM)}\ }\textbf {\bibinfo {volume} {34}},\ \bibinfo
  {pages} {596} (\bibinfo {year} {1987})}\BibitemShut {NoStop}%
\bibitem [{\citenamefont {Thorup}(2000)}]{thorup2000ram}%
  \BibitemOpen
  \bibfield  {author} {\bibinfo {author} {\bibfnamefont {M.}~\bibnamefont
  {Thorup}},\ }\href@noop {} {\bibfield  {journal} {\bibinfo  {journal} {SIAM
  Journal on Computing}\ }\textbf {\bibinfo {volume} {30}},\ \bibinfo {pages}
  {86} (\bibinfo {year} {2000})}\BibitemShut {NoStop}%
\end{thebibliography}
\end{document}